\documentclass[aps,10pt,pre,twocolumn,amsmath,amsfonts,floatfix,superscriptaddress,nolongbibliography]{revtex4-1}
\usepackage{mathrsfs}
\usepackage{graphicx,color}
\usepackage{bm,bbm}
\usepackage{multirow}
\usepackage{hyperref}
\usepackage[caption=false]{subfig}
\usepackage{amsthm,algorithm,algpseudocode}

\usepackage{physics}
\usepackage{mathtools}

\def\OO{\mathcal O}

\newcommand*\bitxor{\wedge}
\newcommand{\tmat}[2]{\prescript{}{\mathrm{#1}}{\mathbf{#2}}}
\DeclareMathOperator\rol{rol}% rotate left
\DeclareMathOperator\ror{ror}% rotate right
\DeclareMathOperator\popc{popc}% popcount
\DeclareMathOperator\netp{netp}% popcount
\newcommand{\perptm}{\prescript{}{\perp}{\mathrm{TM}}}
\newcommand{\paralleltm}{\prescript{}{\parallel}{\mathrm{TM}}}
\newcommand{\diagtm}{\prescript{}{/}{\mathrm{TM}}}
\newcommand{\floor}[1]{\lfloor #1 \rfloor}

\begin{document}

\title{Numerical transfer matrix study of frustrated next-nearest-neighbor Ising models on square lattices}

% authors
\author{Yi Hu}
\email{yi.hu@duke.edu}
\affiliation{Department of Chemistry, Duke University, Durham, North Carolina 27708, USA}
\author{Patrick Charbonneau}
\email{patrick.charbonneau@duke.edu}
\affiliation{Department of Chemistry, Duke University, Durham, North Carolina 27708, USA}
\affiliation{Department of Physics, Duke University, Durham, North Carolina 27708, USA}
\date{\today}

\begin{abstract}
Ising models with frustrated next-nearest-neighbor interactions present a rich morphology of modulated phases. These phases, however, assemble and relax slowly, which hinders their computational study. In two dimensions, strong fluctuations further hamper determining their equilibrium phase behavior from theoretical approximations. 
The exact numerical transfer matrix (TM) method, which bypasses these difficulties, can serve as a benchmark method once its own numerical challenges are surmounted. Building on our recent study [Hu and Charbonneau, Phys. Rev. B \textbf{103}, 094441 (2021)], in which we evaluated the two-dimensional axial next-nearest-neighbor Ising (ANNNI) model with transfer matrices, we here extend the effective usage of the TM method into the Ising models with biaxial, diagonal, and third-nearest-neighbor frustrations (BNNNI, DNNI, and 3NNI models). Thanks to the high-accuracy numerics provided by the TM results, various physical ambiguities about these reference models are resolved and an overview of modulated phase formation is obtained.
\end{abstract}

\maketitle

\section{Introduction}
A ferromagnetic Ising model frustrated by next-nearest-neighbor antiferromagnetic interactions offers a minimal description of systems with short-range attractive and long-range repulsive (SALR) interactions~\cite{sciortino2004equilibrium,zhuang2016recent}. Because even such a simple model family leads a rich set of modulated morphologies---depending on the strength, length scale and orientation of the frustration---it has been used to recapitulate the physics of systems as diverse as magnetic ordering in metals and superconductors~\cite{selke1979monte,dagotto2013colloquium}, and microphase formation in surfactants~\cite{widom1986lattice,dawson1988phase}.

From a statistical physics standpoint, two-dimensional versions of these models are especially interesting. Strong thermal fluctuations alter the nature of phase transitions and lead to novel equilibrium behaviors, such as floating incommensurability and Kosterlitz-Thouless (KT)-type criticality~\cite{kosterlitz1973ordering,pokrovsky1979ground}. These features, however, are challenging to capture in theoretical and numerical studies. (Experimental systems tend to be described by more complex models. See, e.g., Ref.~\cite{glasbrenner2015effect}.)
As a result, long-standing debates persist, including about the putative existence of the critical incommensurate (IC) phase in the axial and biaxial next-nearest-neighbor Ising (ANNNI and BNNNI, respectively) models~\cite{selke1988annni} as well as the order of the phase transition in the diagonal nearest-neighbor Ising (DNNI) model~\cite{moran1993first}. Over the last decade, significant advances have been made in surmounting some of the underlying technical difficulties. For the aforementioned ambiguities in particular, recent studies have confirmed the existence of the IC phase in the ANNNI model~\cite{shirakura2014kosterlitz,matsubara2017domain,hu2021resolving} as well as the order of the transition and the location of the Potts critical point in the DNNI model~\cite{jin2012ashkin,jin2013phase}.
Yet both qualitative and quantitative uncertainties persist, and rather emerge in the light of these advances (see, for instance, Refs.~\cite{bobak2015phase,ramazanov2016thermodynamic,li2021tensor}). A benchmark method that provides exact results for target systems, or with well-controlled limitations would thus be particularly helpful to make complete physical sense of these models.

In this context, the use of numerical transfer matrices (TM), which provide high-accuracy numerical solutions, appears enticing. In fact, the idea is not new. TM were used on frustrated Ising models starting in the 1980s~\cite{pesch1985transfer,beale1985finite,oitmaa1987finite}, but then struggled to provide qualitative---let alone quantitative---insight. The challenge is that only strips of finite width $L$ can be solved with TM, hence a careful finite-size scaling analysis must also be part of the thermodynamic extrapolation, $L\rightarrow\infty$. Given that the algorithmic complexity of TM grows exponentially with $L$ and that frustrated models exhibit large pre-asymptotic corrections, limited computational resources then resulted in physical obfuscation. Exponential improvement to computational hardware over the years (Moore's Law) coupled with more efficient eigensolvers~\cite{lehoucq1998arpack} offer hope that the situation might have since improved. For instance, TM now provide definitive solutions of even fairly complex (quasi-)one-dimensional continuum-space systems~\cite{godfrey2015understanding,robinson2016glasslike,hu2018clustering,hu2018correlation,hu2020comment}. 

In a recent Letter~\cite{hu2021resolving}, we reported the TM resolution of various long-standing ambiguities of the two-dimensional ANNNI model at a reasonable computational cost thanks to the combinations of various algorithmic optimizations. Because this approach is sufficiently generic to be adapted to related lattice models, this article revisits a series of frustrated lattice models: the ANNNI, DNNI, and BNNNI models as well as the generic third-nearest-neighbor Ising (3NNI) model. We notably resolve long-standing ambiguities surrounding the DNNI model at intermediate frustration, and build quantitative phase diagrams for the BNNNI and 3NNI models. The rest of this paper is organized as follows. In Sec.~\ref{sec:m_m}, we provide a complete description of these models, review their key properties, and discuss some of the remaining phase ambiguities, before introducing the TM approach. Section~\ref{sec:r_d} presents results for the various models. A brief conclusion follows in Sec.~\ref{sec:conclusion}.

\section{Models and methods} \label{sec:m_m}

\begin{figure*}	
\includegraphics[width=\textwidth]{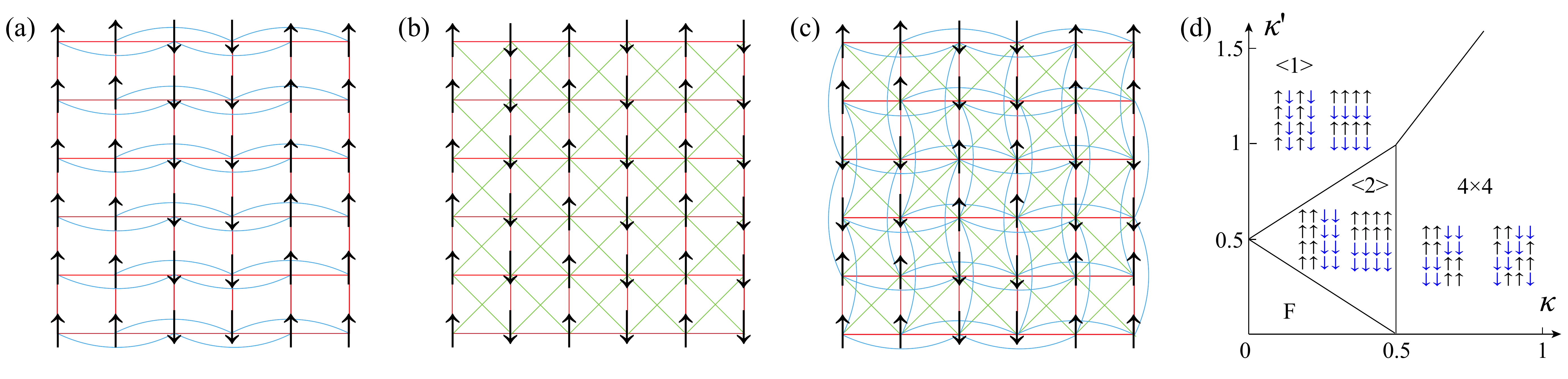}
\caption{Schematics of the (a) ANNNI, (b) DNNI, and (c) 3NNI models on a square lattice. Arrows denote spins, and lines indicate Ising nearest-neighbor (red), diagonal NN (green) and (bi)axial NNN (blue) interactions. (d) $T=0$ phase diagram for the 3NNI model.}
\label{fig:intro-model}
\end{figure*}

In this section we describe the various next-nearest-neighbor Ising models on a square lattice considered in this work, and highlight some of the existing results and predictions. We also briefly describe the numerical TM method. (More details can be found in Appendix~\ref{appd:construction} and~\ref{appd:reducedtmat}.)

\subsection{ANNNI model} \label{sec:model:ANNNI}
The ANNNI model is a minimal model for lamellar microphase formers. Its Hamiltonian reads
\begin{equation} \label{eq:model:ANNNI}
\mathcal{H}_\mathrm{ANNNI} = -J \sum_{\langle i, j \rangle} s_i s_j + \kappa J \sum_{\langle i, j \rangle_\mathrm{ANNN}} s_i s_j - h \sum_i s_i,
\end{equation}
for spin variables $s_i = \pm 1$, coupling constant $J > 0$, next-nearest-neighbor frustration strength along the axial direction $\kappa > 0$ and an external field $h$ (here $J=1$ and $h=0$ for all models considered).
For $\kappa=0$, this model reduces to the standard ferromagnetic Ising model, and for $\kappa < 1/2$, the order-disorder transition remains part of the Ising universality class. For $\kappa > 1/2$, the antiphase $\langle 2\rangle$ (of period 4, $\uparrow \uparrow \downarrow \downarrow$), which forms the energetic ground state, melts into a critical IC phase at $T_\mathrm{c2}$, and then becomes fully disordered at $T_\mathrm{c1}$. Although the existence of the IC phase has long been debated~\cite{selke1988annni}, recent studies provide clear evidence of its persistence, i.e., $T_\mathrm{c1} > T_\mathrm{c2}$ for all $\kappa>1/2$~\cite{shirakura2014kosterlitz,matsubara2017domain,hu2021resolving}.

\subsection{DNNI model} \label{sec:model:DNNI}

The diagonal nearest-neighbor Ising model modifies the Ising model by including non-axial nearest neighbor interactions. Its Hamiltonian reads
\begin{equation} \label{eq:model:DNNI}
\mathcal{H}_\mathrm{DNNI} = -J \sum_{\langle i, j \rangle} s_i s_j + \kappa J \sum_{\langle i, j \rangle_\mathrm{DNN}} s_i s_j - h \sum_i s_i,
\end{equation}
using the same parameter notation as for the ANNNI model. (This model is also known as the next-nearest-neighbor or frustrated Ising model and as the $J_1$-$J_2$ Ising model, but we here denote it DNNI to distinguish it more saliently from the other models considered.) 
The energetic ground state of the DNNI model is ferromagnetic for $\kappa < 1/2$ and striped with $\langle 1 \rangle$ (of period 2, $\uparrow \downarrow$) for $\kappa > 1/2$. In both cases, melting leads directly to a disordered phase, but whether or not the transition remains first-order in nature around $\kappa = 1/2$ has long been debated. The putative first-order regime indeed appears to narrow as the quality of numerical studies improves.
Predictions for a Potts point dividing the weakly-first-order from the continuous transition vary from $\kappa^* \simeq 0.9$~\cite{dos2008phase,kalz2011analysis}, $0.67(1)$~\cite{jin2012ashkin,jin2013phase} to $\simeq 0.54$~\cite{li2021tensor}.
For $\kappa < 1/2$, it has generally been assumed that the transition is part of the Ising universality class, but recent cluster mean-field approximation and effective-field theory study suggest that a narrow first-order transition regime might also be found at $\kappa \lesssim 1/2$~\cite{jin2013phase,bobak2015phase}.
Several simulation studies further suggest a nonzero transition temperature at $\kappa=1/2$ proper, i.e., $T_c\mathrm(\kappa=1/2) > 0$~\cite{boughaleb2010bicritical,ramazanov2016thermodynamic,timmons2018role}, in marked contrast to prior works~\cite{landau1980phase,moran1993first,kalz2008phase,kalz2009monte,kim2010partition}. Because theoretical approximations behave irregularly around $\kappa \lesssim 1/2$~\cite{oitmaa1981square}, however, a conclusive assessment has thus far remained out of reach.

\subsection{BNNNI and 3NNI models} \label{sec:model:3NN}

Including biaxial next-nearest-neighbor interactions, i.e., couplings with Euclidean third-nearest-neighbor, to the DNNI model gives rise to the 3NNI model (also known as the $J_1$-$J_2$-$J_3$ Ising model~\cite{liu2016role}). Its Hamiltonian reads
\begin{equation} \label{eq:model:3NN}
\begin{aligned}
\mathcal{H}_\mathrm{3NNI} = &-J \sum_{\langle i, j \rangle} s_i s_j + \kappa J \!\!\!\!\!\!\sum_{\langle i, j \rangle_\mathrm{BNNN}}\!\!\!\!\!\! s_i s_j + \kappa' J\!\!\!\! \sum_{\langle i, j \rangle_\mathrm{DNN}} \!\!\!\!\!\! s_i s_j \\
&- h \sum_i s_i,
\end{aligned} \end{equation}
and setting $\kappa' = 0$ recovers the BNNNI model, which we consider first.

The BNNNI model has a ferromagnetic ground state for $\kappa < 1/2$, and the transition to the high-temperature paramagnetic phase is thought to exhibit Ising universality in that regime. For $\kappa > 1/2$ the model presents two distinct energetic ground states [Fig.~\ref{fig:intro-model}(d)]: $4 \times 4$ checkerboard order, or diagonal stripes of width 2. Early Monte Carlo simulations suggested that melting of these structures proceeds through a first-order transition~\cite{landau1985phase}, but later work found the thermodynamic evolution to be more complicated. Multiple metastable states indeed develop at intermediate temperatures~\cite{velgakis1988critical}, and a two-step transition involving a critical IC phase at $1/2 < \kappa < \kappa^*$~\cite{oitmaa1987finite,aydin1989renormalisation,aydin1989monte,dasgupta1991bnnni} has been proposed. It is further unclear whether the Lifshitz point takes place at finite $\kappa^*$, or whether $\kappa^* \rightarrow \infty$. Even studies suggesting the former offer but a qualitative determination of $\kappa^*$~\cite{aydin1989monte}. 

Landau and Binder determined the energetic ground state structure of the more general 3NNI model with antiferromagnetic Ising interactions~\cite{landau1985phase} (see Fig.~\ref{fig:intro-model}(d)), which can be mapped onto ferromagnetic Ising interactions by flipping every other spin on the lattice. In short, while the ferromagnetic (F) phase persists for $\kappa + \kappa' < 1/2$, the ground state either follows that of the BNNNI model---$4 \times 4$ checkerboard or diagonal stripes of width 2 (both denoted $4 \times 4 $, for convenience)---at large $\kappa$, or that of the DNNI model---$\langle 1\rangle$ stripes---at large $\kappa'$. At large frustration, the separation line is given by $\kappa' = 2\kappa$; and at intermediate frustration, the $\langle 2 \rangle$ phase is also a ground state. By contrast to the ANNNI model, however, the modulation can here grow along either axial directions (on a square lattice), hence the ground state is eight-fold (instead of four-fold) degenerate. The transition is further thought to be first-order---as for the 8-state Potts model---instead of continuous~\cite{liu2016role}. 

Interestingly, the lattice gas representation of the 3NNI model is also the two-dimensional counterpart to the Widom-Wheeler lattice microemulsion model~\cite{dawson1988phase} (with $h$ corresponding to the chemical potential). The 3NNI model presents a $\langle 2 \rangle$ phase---a lamellar microphase in the language of Ref.~\onlinecite{dawson1988phase}---as a result of the competition between DNN and BNNN interactions. We thus here concentrate on this particular regime for the investigation of the 3NNI model.

\subsection{Numerical TM method} \label{sec:method:tm}
Although these two-dimensional models lack an analytic solution, TM can numerically solve the exact partition function of a semi-infinite strip of width $L$, and the results can then be analyzed using finite-size scaling approaches to extrapolate the thermodynamic behavior in the limit $L \rightarrow \infty$. 

Generically, TM encode the interaction between subsequent layers states $a$ and $a'$ as 
\begin{equation} \label{eq:tmatgeneral}
\mathbf{T}_{a,a'} = \exp[-\beta (V_x(a) + V_z (a, a') ) ],
 \end{equation}
where $\beta=1/k_B T$ is the inverse temperature (the Boltzmann constant is set to unity, $k_B=1$), and $V_x$ and $V_z$ are intra- and inter-layer interaction energies, respectively. The partition function of the strip of length $N$ is given by $\tr( \mathbf{T}^N )$, and
in the limit of $N \rightarrow \infty$, $Z$ is given by the leading eigenvalue of $\mathbf{T}^N$, $\lambda_0^N$. The free energy per spin is thus
\begin{equation} \label{eq:freeEnergy}
\beta f = -\frac{\ln \lambda_0}{L},
\end{equation}
and the marginal probability of a state $a$ is given by the product of (normalized) left and right eigenvector of $\lambda_0$, $P(a) = \varphi_0^{-1}(a) \varphi_0 (a)$ (after normalizing as $\sum_a P(a) = 1$). Given the leading eigenvalue and eigenvector, thermodynamic observables can be obtained exactly, including the internal energy per spin, $\beta u=-\partial (\beta f) / \partial T$, and the specific heat per spin, $c = \partial u/\partial T$. 

One of the key advantages of TM is that they also provide the correlation length $\xi$, which help identify the location and nature of phase transitions. Given that the conditional probability of finding a subsequent layer state $a'$ after layer $a$ is~\cite{hu2020comment}
\begin{equation}
P(a'|a) = \frac{ \mathbf{T}_{a,a'} \varphi_0 (a') }{\lambda_0 \varphi_0 (a) },
\end{equation}
the generic conditional probability for $a'$ having a distance $N$ from $a$ is then
\begin{equation} \begin{aligned}
P^{(N)}(a'|a) &= \frac{ \mathbf{T}^N_{a,a'} \varphi_0 (a') }{\lambda_0^N \varphi_0 (a) } = P(a') + \\ 
&\sum_{i=1}^{|\{a\}| -1 } \left( \frac{ \lambda_i }{\lambda_0} \right)^N \frac{\varphi_i (a) \varphi_i^{-1} (a') \varphi_0 (a') }{\varphi_0 (a) }, \\
P^{(N)}(a'|a) - P(a') &\sim \exp[N \ln( \frac{|\lambda_1| }{ \lambda_0 }) ] \\
&= \exp(-N/\xi_1 ),
\end{aligned} \end{equation}
where $|\{a\}|$ is the total number of states and $\xi_1 = -1/\ln(|\lambda_1| / \lambda_0)$ is the leading correlation length. Subdominant lengths can be analogously defined as $\xi_i= -1/\ln(|\lambda_i| / \lambda_0)$ for $i>1$.

Beyond that point, each model presents certain peculiarities. (TM prescripts are thus used to distinguish between the ANNNI (A), BNNNI (B), DNNI (D) and 3NNI (3) models as well as to denote the direction of propagation.) For the ANNNI model, because the interaction is anisotropic, the TM can be propagated either perpendicularly ($\tmat{\perp,\mathrm{A}}{\mathrm{TM}}$)~\cite{pesch1985transfer} or parallel ($\tmat{\parallel,\mathrm{A}}{\mathrm{TM}}$)~\cite{beale1985finite} to the direction of the next-nearest-neighbor interaction. Although the former is markedly smaller $2^L \times 2^L$ (vs $4^L \times 4^L$), in the modulated regime the latter converges much faster to the asymptotic scaling as $L$ increases. For the DNNI model, because interactions reach no further than the first subsequent layer, $\tmat{\perp,D}{\mathrm{TM}}$ can be constructed by modifying $\tmat{\perp,\mathrm{A}}{\mathrm{TM}}$ and is thus also of size $2^L \times 2^L$. 
For the BNNNI and 3NNI models, TM propagation along the diagonal of the square lattice, $\diagtm$, has been suggested as preferable at finite $L$~\cite{oitmaa1987finite}, in order to better capture the ordering in the direction of the antiphase modulation. This choice also results in a TM of size $2^L \times 2^L$. In all cases, the TM size can be significantly reduced by leveraging the model symmetry, as described in Appendix~\ref{appd:reducedtmat}. (In practice, the TM is not explicitly computed but implicitly represented by a matrix-vector subroutine, as described in Appendix~\ref{appd:construction}).)

\section{Results and discussion} \label{sec:r_d}

In this section we present TM results for various frustrated models at finite $L$ as well as the finite-size scaling analysis used to extrapolate the thermodynamic behavior in the limit $L \rightarrow \infty$. We provide results for the ANNNI model that complement our recent analysis of that system~\cite{hu2021resolving}, and discuss the behavior of the DNNI model, paying particular attention to the physical ambiguities previously reported in the literature. For the more computationally challenging BNNNI model we propose a phase diagram and discuss various remaining uncertainties. Finally, we examine the $\langle 2 \rangle$ regime of the generic 3NNI model with $\kappa = \kappa'/2$, which are frustration conditions akin to those of a three-dimensional microphase former.

\subsection{ANNNI model}
\label{sec:annniresults}

For the ANNNI model, we first obtain the subleading correlation length using both $\perptm$ and $\paralleltm$. These results mainly serve as reference for other models. In order to probe further the existence of a critical IC (or floating) phase, we also compare the domain wall free energy, using a scheme first proposed for the density matrix renormalization group (DMRG) approach~\cite{derian2006modulation}.

\begin{figure}
\includegraphics[width=0.98\columnwidth]{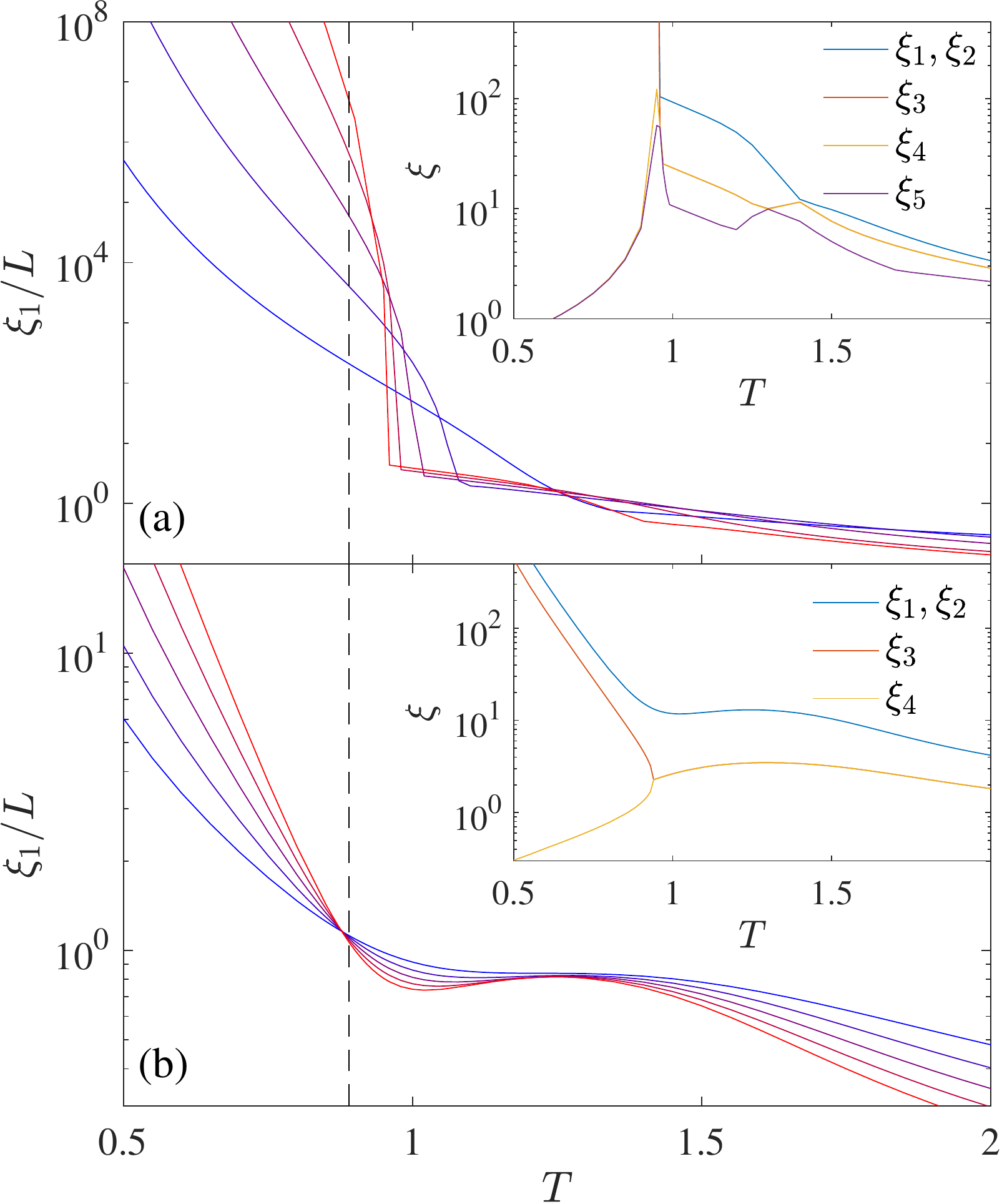}
\caption{Leading correlation length for the ANNNI model with $\kappa=0.6$ in (a) $\perptm$, $L=8, 12, ..., 24$ and (b) $\paralleltm$, $L=8, 10, ..., 16$, from blue to red. The ordering transition position, $T_\mathrm{c2} = 0.89(1)$ (dashed line) is given as reference. Results from $\paralleltm$ converge more smoothly and cleanly to the thermodynamic limit. Inset: First few leading correlation lengths in (a) $\perptm$ for $L=24$ and (b) $\paralleltm$ for $L=16$. Eigenvalue crossings in the former leads to a complex $T$ dependence of the corresponding $\xi_1$.}
\label{fig:ANNNIclength}
\end{figure}

Comparing the leading correlation length, $\xi_1$, obtained from $\perptm$ and $\paralleltm$ highlights the anisotropic nature of the model (Fig.~\ref{fig:ANNNIclength}). 
In particular, as $T$ increases results from $\perptm$ decay non-smoothly due to the crossing of sub-dominant correlation lengths [Fig.~\ref{fig:ANNNIclength}(a), inset]. This phenomenon, which generically accompanies a structural crossover~\cite{hu2018correlation}, here bespeaks a stepwise change in the modulation period~\cite{hu2021resolving}. 
For $L=24$, for example, two distinct steps can be identified, both involving $\xi_1=\xi_2$ (associated with doubly degenerate eigenvalues) crossing the subleading $\xi_3$ and $\xi_4$. These features, however, complicate the evolution of $\xi_1$, and hence hinder the local exponent analysis and the identification of the IC phase~\cite{beale1985finite,oitmaa1987finite,hu2021resolving}. By contrast, results for $\paralleltm$ evolve smoothly with $T$. The non-monotonic growth of $\xi_1=\xi_2$ (associated with complex conjugate eigenvalues) at intermediate $T$ can thus be construed as a signature of the critical IC phase, and its boundaries, $T_\mathrm{c1} > T_\mathrm{c2}$, can be identified by analyzing the local exponent~\cite{hu2021resolving}. In addition, the subleading correlation lengths $\xi_3$ and $\xi_4$ are found to merge at the $\langle 2 \rangle$-to-IC phase transition temperature, $T_\mathrm{c2}(L)$, hence providing an estimate of $T_\mathrm{c2}$ that is fully consistent with those from Ref.~\onlinecite{hu2021resolving}.

\begin{figure}
\includegraphics[width=0.98\columnwidth]{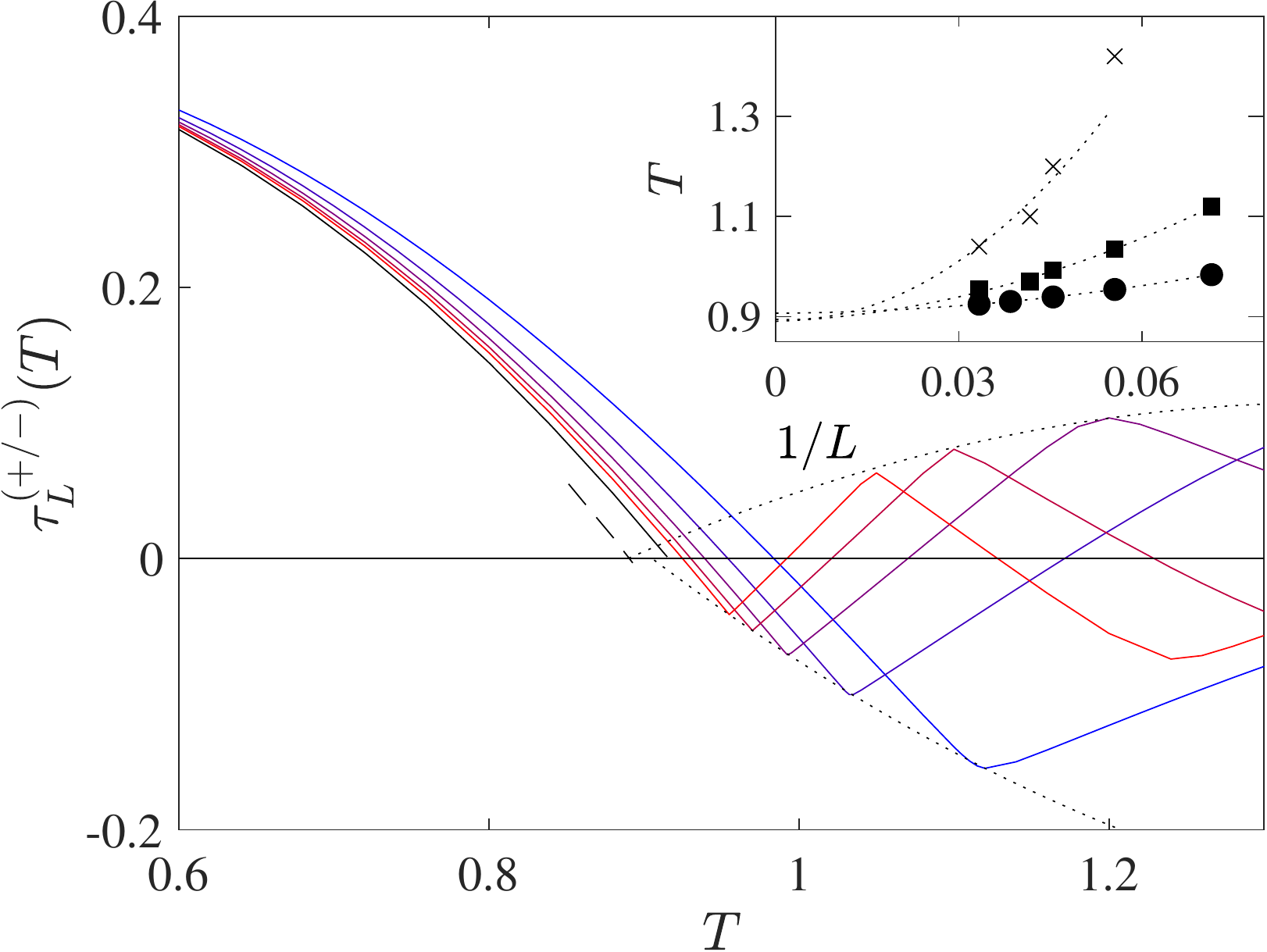}
\caption{Domain wall free energy $\tau^{(+/-)}_L(T)$ for the ANNNI model with $\kappa=0.6$ from $\perptm$ with $L=14,18,...,28$ from blue to red lines. The extrapolation of $\tau^{(+/-)}(T)$ from Eq.~\eqref{eq:dwfextrapolate} with the scaling expected for the antiphase, $B=2$ (black solid line), vanishes at $T=0.91(1)$. The extrapolation with the scaling expected for the critical point, $B=1$ (dashed black line), gives $T_\mathrm{c2}=0.89(1)$. Both are numerically consistent with previous estimates. Inset: extrapolating the positions of local peaks (crosses), valleys (squares) and the first zeros (circles) are consistent with $T_\mathrm{c2}$ (dotted lines are guides for the eye).
}
\label{fig:ANNNIdomain}
\end{figure}

The stepwise change in the modulation period can also been identified from the domain-wall free energy obtained from comparing two systems under different boundary conditions~\cite{richards1993numerical},
\begin{equation} \label{eq:dwfenergy}
\tau^{(+/-)}_L(T) = L[f_L^{(+/-)}(T) - f_L^{(+/+)}(T) ] \sigma_L ,
\end{equation}
where $f_L^{(+/-)}$ (or $f_L^{(+/+)}$) are the free energy of a system fixing $s_1 = +1$ and $s_L = -1$ (or $+1$), respectively, and $\sigma_L = \pm 1$ in that the ground state $\tau^{(+/-)}_L$ is positive. In the $\langle 2 \rangle$ phase, and for modulations congruent with $L \mod 4 \equiv 2$, the expected finite-size scaling is~\cite{richards1993numerical}
\begin{equation} \label{eq:dwfextrapolate}
\tau^{(+/-)}(T) - \tau^{(+/-)}_L(T) \sim L^{-B},
\end{equation}
where the exponent is observed to be $B=2$ in the $\langle 2\rangle$ phase~\cite{richards1993numerical}, but pre-asymptotic corrections grow upon approaching the critical temperature, $T_\mathrm{c}$, whereat scaling theory gives $B=1$~\cite{privman1990finite}.

Following Ref.~\onlinecite{derian2006modulation}, we first fit the results with $B=2$ and extrapolate the thermodynamic $\tau^{(+/-)}(T)$ over a broad range of $0.6 < T \lesssim T_\mathrm{c2}$. The result is fully consistent with that former study, predicting $T_\mathrm{c2}=0.91(1)$ (vs 0.907~\cite{derian2006modulation}).
Alternatively, setting $B=1$ (the expected critical scaling) gives $T_\mathrm{c2}=0.89(1)$. The two estimates thus differ only marginally, and are both consistent with recent quantitative estimates for the transition temperature~\cite{matsubara2017domain,hu2021resolving}.

For $T > T_\mathrm{c2}$, $\tau^{(+/-)}_L(T)$ oscillates around $0$, suggesting that the modulation period varies with $T$, a clear signature of the IC phase. Extrapolating the first peak and valley further suggests that these oscillations coalesce at $T_\mathrm{c2}$ and thus vanish in the thermodynamic limit $L \rightarrow \infty$. The zigzagging behavior weakens at larger $\kappa$ yet the smoother oscillations around $0$ are still observable at large $L$ for $\kappa \gtrsim 1.5$. Such oscillatory yet vanishing free energy difference for the ($+/+$) and ($+/-$) boundary conditions supports the floating IC phase scenario. The interfacial results are therefore fully consistent with the length scale analysis of Ref.~\onlinecite{hu2021resolving}, and are also consistent with the DMRG results~\cite{derian2006modulation} obtained for much larger systems ($L \sim 10^2$ vs $10^1$ in TM).

\subsection{DNNI model}

For the DNNI model, we first compare the correlation length, internal energy and domain wall free energy results with those of the ANNNI model from Ref.~\onlinecite{hu2021resolving} and Sec.~\ref{sec:annniresults}. We then attempt to resolve the phase ambiguities described in Sec.~\ref{sec:model:DNNI}. 

\subsubsection{Overview of thermodynamic observables}

\begin{figure*}[t!]
\includegraphics[width=\textwidth]{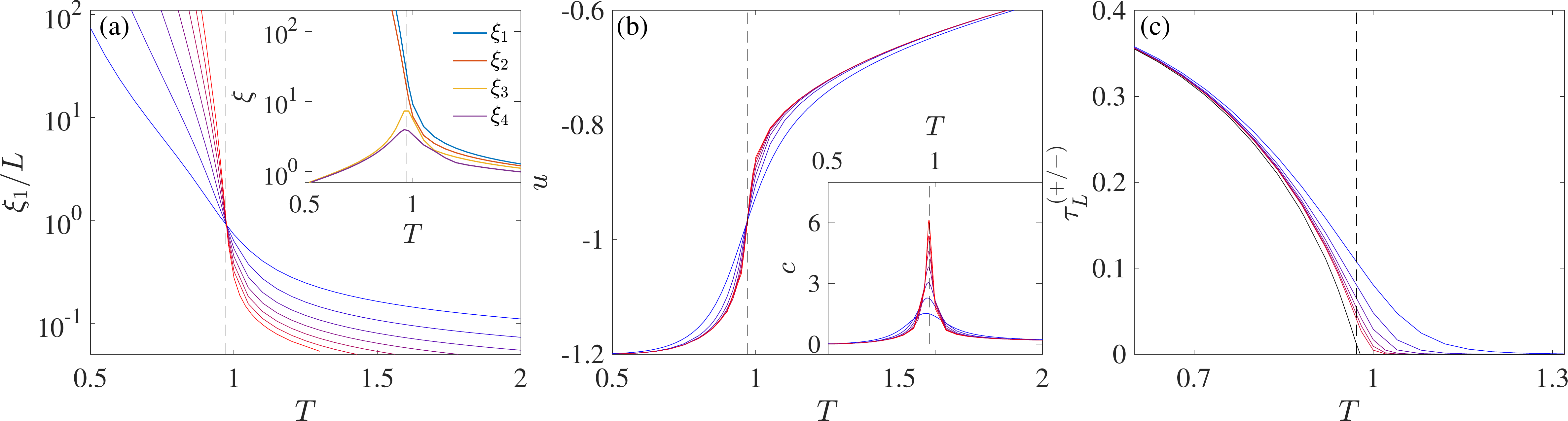}
\caption{Finite-size results for various observables of the DNNI model with $\kappa=0.6$ for $L=8, 12, ..., 32$ (from blue to red lines). Recall that $T_\mathrm{c} = 0.971(1)$ (dashed vertical lines). (a) The leading correlation length cross at single fixed point for all $L$. Inset: the first few leading correlation lengths for $L=24$. No correlation length splitting or crossing is observed. (b) Energy per spin $u$ and (Inset) specific heat $c$ per spin. Both $\xi_1/L$ and $u$ present a fixed point and $c$ peaks sharply at $T_\mathrm{c}$.
(c) Domain wall free energy for $L=14, 18, ..., 28$ from blue to red lines. The extrapolated $\tau^{(+/-)}$ from Eq.~\eqref{eq:dwfextrapolate} (black solid line) vanishes at $T_\mathrm{c}$ (dashed line). 
}
\label{fig:DNNIthermo}
\end{figure*}

The evolution of $\xi_1$ for the DNNI model is smooth and monotonic (see, e.g., Fig.~\ref{fig:DNNIthermo}(a)); no eigenvalue crossing or splitting is observed. 
Unlike the ANNNI model, which presents an algebraic growth, $\xi_1 \sim L^\theta(T)$, over a temperature range~\cite{beale1985finite,hu2021resolving}, 
the DNNI model displays an algebraic scaling only at a single temperature. Also, the anisotropy exponent is then $\theta=1$ ($\xi_1/L \approx \mathrm{const}$), as expected for models with isotropic interactions~\cite{nightingale1976scaling}. 
The crossing point of $\xi_1/L$ further provides an accurate estimate of $T_\mathrm{c}$. Other robust estimators include the crossing point of the internal energy and the peak of the specific heat [Fig.~\ref{fig:DNNIthermo}(b)]. Because the finite-$L$ transition temperature $T^*(L)$ identified by these estimators changes little with $L$, transition estimates can often be identified with up to five significant digits~\cite{jin2013phase}. (Around $\kappa =1/2$, the situation is more complex, as discussed below.)

We also evaluate the domain wall free energy using Eq.~\eqref{eq:dwfenergy} [Fig.~\ref{fig:DNNIthermo}(c)]. As expected, $\tau^{(+/-)}_L$ follows the finite-size scaling of Eq.~\eqref{eq:dwfextrapolate} in the ordered striped phase. For $B=1$, it is extrapolated to vanish at $T_\mathrm{c}$. (The $\sim0.01$ deviation from $T_\mathrm{c}$ is likely due to pre-asymptotic corrections.)
Unlike in Fig.~\ref{fig:ANNNIdomain}, here no signature of oscillation is observed for $\tau^{(+/-)}_L(T > T_\mathrm{c})$. This monotonic evolution is robust for various $\kappa$ both below and above $1/2$. The incommensurate phase---or its finite $L$ echo---is thus clearly absent in the DNNI model.

\subsubsection{$T_\mathrm{c}$ determination}

\begin{figure}
\includegraphics[width=1\columnwidth]{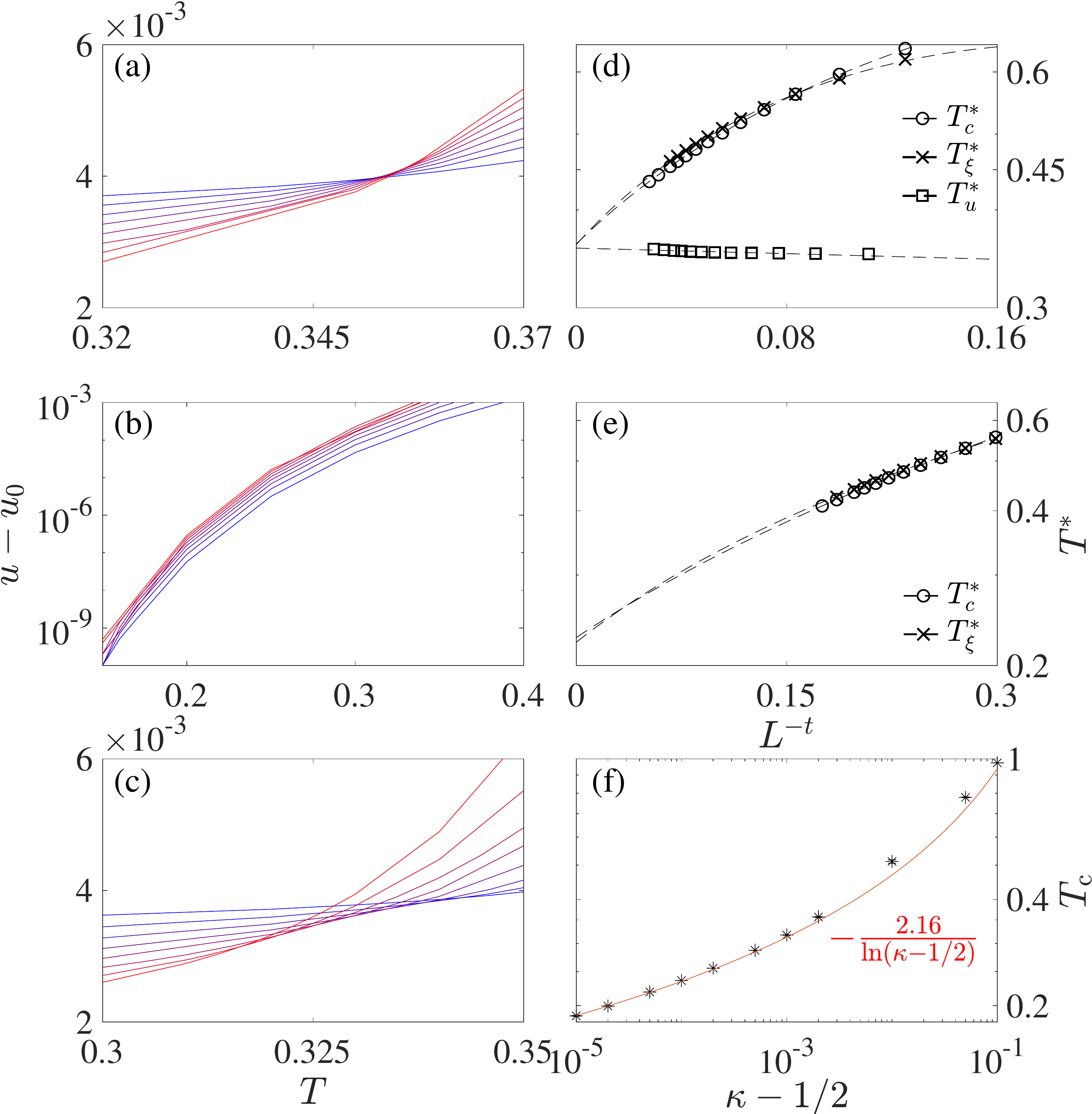}
\caption{Phase transition determination around $\kappa=1/2$. (a-c) Internal energy above the ground state, $u_0=-(1 + 2|\epsilon|)$, for $\epsilon=\kappa-1/2=0.002, 0, -0.002$. (d, e) extrapolation of $T_\mathrm{c}$ via different estimators indicated in legends (see text for detail) for $\epsilon=0.002$ and $0$, respectively. Dashed lines are fitted using Eq.~\eqref{eq:dnniextrapolateT} with empirical exponent $t=1$ and $0.486$, respectively (quadratic fits are used for $T^*_c(L)$ and $T^*_{\xi}(L)$). The results from different estimators are mutually consistent; and for (e) it reproduces Ref.~\onlinecite{ramazanov2016thermodynamic}. (f) $T_\mathrm{c}$ vanishes logarithmically as $\epsilon\rightarrow 0^+$.}
\label{fig:DNNImpext}
\end{figure}

With an understanding of the behavior for various phase transition estimators in hand, we now quantitatively evaluate $T_\mathrm{c}$. In particular, we wish to determine whether $\lim_{\epsilon\rightarrow 0}T_\mathrm{c}(1/2+\epsilon) =0$.
As shown in Fig.~\ref{fig:DNNImpext}(a,c), a crossing point in $T$-$u$ curve is detected for both side of $\kappa=1/2 \pm 0.002$, but is absent right at $\kappa=1/2$ down to numerical accuracy (in practice, $\sim 10^{-10}$) [Fig.~\ref{fig:DNNImpext}(b)]. The crossing thus seemingly takes place at $T=0$, as does the phase transition, but further evidence is needed.

A systematic comparison reveals that for $\epsilon>0$ $T^*_u(L)$ barely shifts with $L$, while for $\epsilon<0$ significant pre-asymptotic corrections appear [see Fig.~\ref{fig:DNNImpext}(a,c)]. We thus apply an empirical fitting form to extrapolate $T_\mathrm{c}$,
\begin{equation} \label{eq:dnniextrapolateT}
T^*(L) = T_c + A L^{-t}
\end{equation}
with a fitting constant $A$ and empirical exponent $t$ ($t=1$ for $\kappa>1/2$). 
Note that other estimators, such as the location of specific heat and the correlation length peaks [see Fig.~\ref{fig:DNNIthermoq}(c)], $T^*_c(L)$ and $T^*_{\xi}(L)$, provide consistent estimates [Fig.~\ref{fig:DNNImpext}(d)], but require a quadratic correction, $B L^{-2t}$, to Eq.~\eqref{eq:dnniextrapolateT} and are thus less accurate. 
Thanks to the exceptionally small finite-size corrections to $T^*_u(L)$ and the high TM accuracy, $T_\mathrm{c}$ can be determined down to $\epsilon=10^{-5}$ [Fig.~\ref{fig:DNNImpext}(f)]. (For $\epsilon<0$, the shift of $T^*_u$ with $L$ makes a comparable extrapolation more haphazard.) Remarkably, for $\epsilon>0$ the resulting transition temperature scales logarithmically as
\begin{equation}
T_\mathrm{c}(\kappa) \approx -\frac{2.16}{\ln (\kappa-1/2)}.
\end{equation}
We thus confidently conclude that $T_\mathrm{c}(\kappa=1/2) = 0$. 

To better understand why Ref.~\onlinecite{ramazanov2016thermodynamic} concluded differently, we replicate their analysis in Fig.~\ref{fig:DNNImpext}(e) (with the same $t=0.486$), and consider $T^*_{\xi}(L)$ as well. Both extrapolations give $T_c(\kappa=1/2)=0.22(1)$, as Ref.~\onlinecite{ramazanov2016thermodynamic} found. Previous extrapolation attempts have thus been obfuscated by the complex and significant pre-asymptotic corrections to various variables around $\kappa=1/2$, as we discuss in Sec.~\ref{sec:dnni1storder}.

\subsubsection{Order of transition}
\label{sec:dnni1storder}
\begin{figure}
\includegraphics[width=0.98\columnwidth]{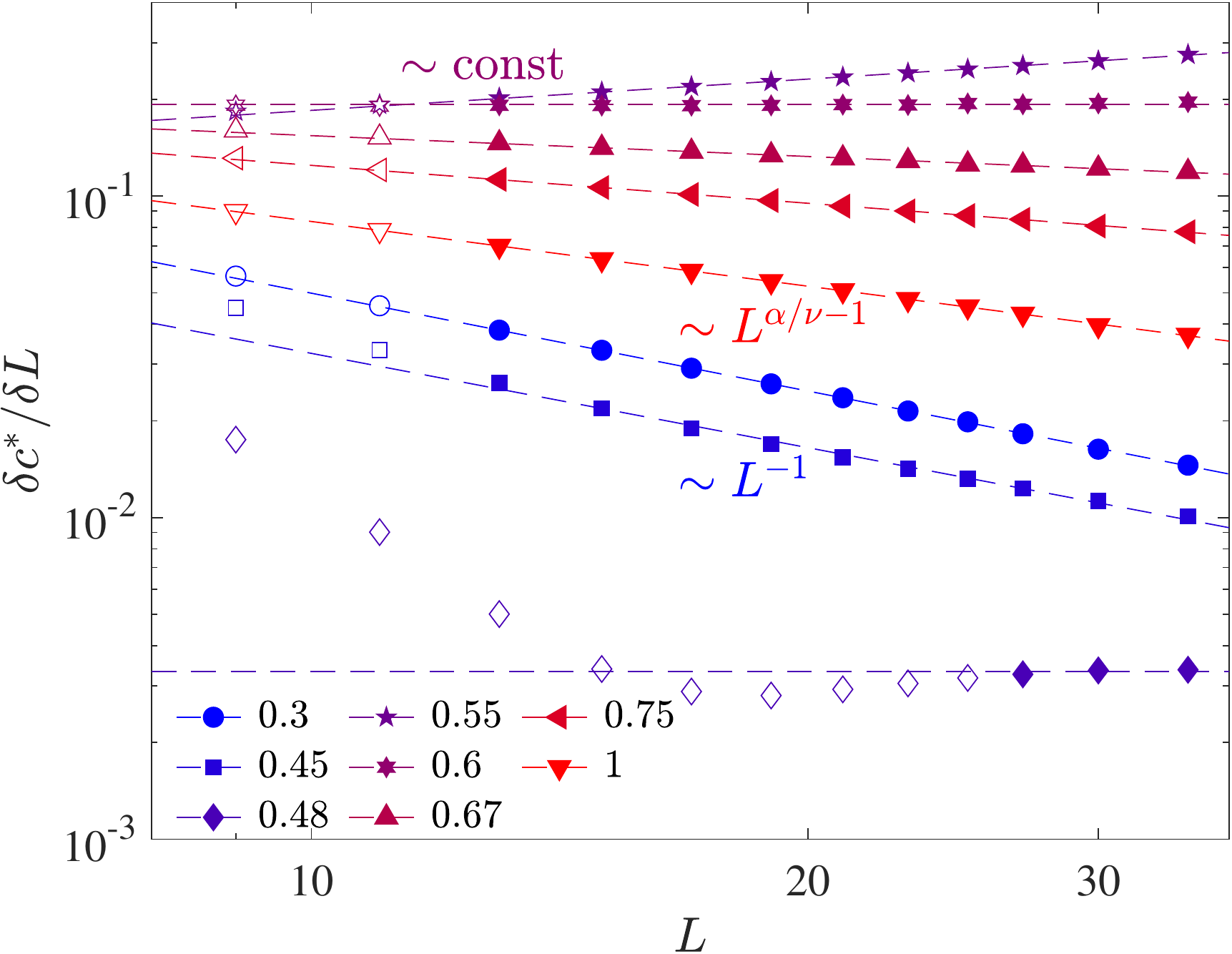}
\caption{Finite-size scaling of the DNNI peak specific heat for various $\kappa$. $\delta c^*/\delta L \sim L^{-1}$ for the Ising-type continuous transition ($\kappa=0.3,0.45$); $\sim L^{\alpha/\nu-1}$ with $0 < \alpha/\nu \le 1$ for the AT-type continuous transition ($\kappa=0.67,0.75,1$) and is expected to approach constant for the first-order scenario ($\kappa=0.48,0.55,0.6$) but the slope of the fitting line varies continuously in $0.5 < \kappa < \kappa^*$.}
\label{fig:DNNItorder}
\end{figure}

\begin{figure}
\includegraphics[width=0.98\columnwidth]{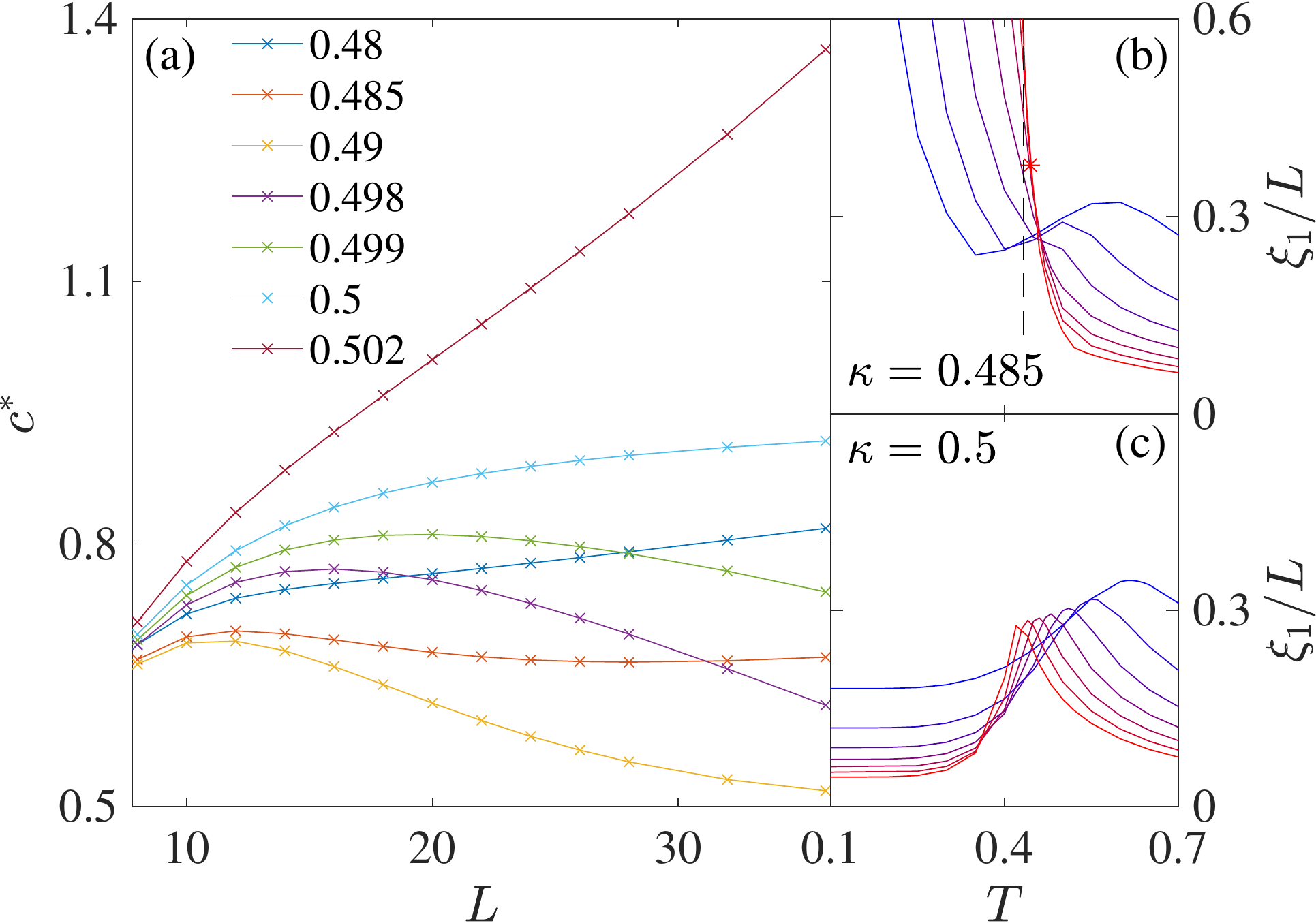}
\caption{The DNNI model exhibits pronounced pre-asymptotic corrections around $\kappa = 1/2$: (a) peak specific heat $c^*(L)$ for various $\kappa$ around $1/2$; and (b, c) leading correlation length for $L=8, 12, ..., 32$ at $\kappa=0.485$ and $1/2$ with $T_\mathrm{c}$ (dashed line in b). For $\kappa=0.485$, as $L$ grows the crossing points between subsequent $\xi_1/L$ (asterisk between $L=28$ and $32$) approach $T_\mathrm{c}$, while the local peak vanishes.
 }
\label{fig:DNNIthermoq}
\end{figure}

Another actively debated aspect of the DNNI model is the order of its various ordering phase transitions. In principle, this can be determined from the scaling of the peak specific heat:
\begin{enumerate}
\item For an Ising-type continuous phase transition (expected for small $\kappa$),
\begin{equation}
c^*(L) \approx A \ln L + c_0,
\end{equation}
$c_0$ denotes the \emph{background} specific heat and $A$ is a fitting constant. 

\item For a first-order transition (expected for $1/2 < \kappa < \kappa^*$ and speculated for $\kappa^\dagger<\kappa<1/2$), 
\begin{equation}
c^*(L) = A L + c_0.
\end{equation}

\item For an Ashkin-Teller (AT)-type phase transition (expected for $\kappa \ge \kappa^*$), 
\begin{equation}
c^*(L) = A L^{\alpha/\nu} + c_0,
\end{equation}
where $\alpha$ is the heat capacity exponent and $\nu$ is the correlation length exponent, and $0 < \alpha/\nu \le 1$. In particular, $\alpha/\nu = 1$ characterizes the Potts critical point.

\end{enumerate}
To eliminate the background correction, we here consider the finite differentiate, which scales as 
\begin{equation}
\label{eq:alphafit}
\delta c^*/\delta L = [c^*(L+1) - c^*(L - 1) ]/2 = A L^b,
\end{equation}
with $b = \alpha/\nu - 1$ for continuous transitions and with $b=0$ (a plateau) for first-order transitions. Results for selected $\kappa$ are reported in Fig.~\ref{fig:DNNItorder}. Fitting Eq.~\eqref{eq:alphafit} gives $\alpha/\nu=-0.01(3)$ and $0.09(10)$ for $\kappa=0.3$ and $0.45$, respectively, both consistent with an Ising-type transition with $\alpha=0$.
In the (expected) weakly first-order regime $1/2 < \kappa < \kappa^*$, however, the fitting slope decreases with $\kappa$. For instance, $\kappa=0.55$ and $ 0.6$ give $b = 0.32(2)$ and $0.02(1)$, respectively, instead of $b=0$ throughout. This drift, which was also reported in Monte Carlo simulations of small systems~\cite{landau1985phase}, suggests that pronounced finite-size corrections are a play. The TM approach thus still cannot clearly identify $\kappa^*$. 
Nevertheless, the regime of effective $\alpha/\nu > 1$ deviates sufficiently significantly from the AT scenario ($0 < \alpha/\nu \le 1$) to marginally favor the weakly first-order over the continuous AT-type transition. By contrast, the drift of $\alpha/\nu$ observed at larger $\kappa$ (e.g., $\kappa=0.75$ and $1$ give $\alpha/\nu = 0.61(2)$ and $0.33(2)$, respectively) is consistent with the AT-type transition with varying exponent~\cite{jin2012ashkin,jin2013phase}.

By contrast, the Ising-type (at small $\kappa$) and the weakly first-order (speculated for $\kappa^\dagger<\kappa < 1/2$) regimes can be distinguished more straightforwardly from $b=-1$ and $0$, respectively. 
A clear $b=-1$ scaling persists at least up to $\kappa = 0.45$. For $\kappa=0.48$, however, $\delta c^*/\delta L$ first decreases with $L$, and then grows slightly before plateauing. Although pre-asymptotic features partly muddle the physical picture, this trend clearly deviates from physical expectations for an Ising-type transition. It is instead reminiscent of a weakly first-order transition, and thus support the theoretical speculations of Refs.~\onlinecite{jin2013phase,bobak2015phase} that such a regime should exist for $\kappa^\dagger<\kappa<1/2$ with $\kappa^\dagger\gtrsim 0.45$. (Reference~\onlinecite{ramazanov2016thermodynamic} concluded that a continuous transition takes place for $\kappa=0.48$, based on the absence of discontinuity in $u(T)$, but did not consider the weakly first-order transition scenario.) 
Interestingly, a recent computation for a low-connectivity Bethe lattice suggests that such a weakly first-order transition in the vicinity of the multicritical point ($\kappa \lesssim 1/2$ in the DNNI model) is a mean-field feature~\cite{charbonneau2021solution}. This behavior thus clearly differs from the fluctuation-induced discontinuity of the weakly first-order transition for $1/2<\kappa<\kappa^*$.

The difference might also explain the markedly distinct scaling properties observed on either side of $\kappa \rightarrow 1/2$ [as noted in Fig.~\ref{fig:DNNImpext}(a,c)]. 
Slightly above $\kappa =1/2$, even thought pre-asymptotic corrections prevent a quantitative determination of $\alpha/\nu$ (see, e.g., $\kappa=0.55$ in Fig.~\ref{fig:DNNItorder}), the monotonic growth of $c^*(L)$ is qualitatively consistent with a weakly first-order scenario. For example, at $\kappa=0.502$, $c^*(L)$ grows nearly linearly already for $L \ge 28$ [Fig.~\ref{fig:DNNIthermoq}(a)]. 
By contrast, slightly below $\kappa=1/2$ deviations from scaling are confounding even for the sake of qualitative speculations. For instance, for $\kappa = 0.485$, $c^*(L)$ decreases at intermediate $L$ before increasing again. 
From $\kappa=0.485$ to $0.499$, this pre-asymptotic behavior extends to even larger $L$ as $\kappa\rightarrow1/2$. Moreover, the range of small $L$ growth extends as well, leaving but a purely monotonic growth at $\kappa = 1/2$. The evolution of $\xi_1$ also hints at a complex finite-size behavior for $\kappa \lesssim 1/2$. For example, for $\kappa=0.485$, a local peak appears at small $L$ but disappears as $L$ increases [Fig.~\ref{fig:DNNIthermoq}(b)], and then a crossing is recovered around $T_\mathrm{c}$. 
As $\kappa$ further approaches $1/2$, this local peak survives for larger systems and is expected to persist for all $L\rightarrow\infty$ at $\kappa=1/2$ [Fig.~\ref{fig:DNNIthermoq}(c)]. In this limit case, $\xi_1$ approaches a constant at both low and high $T$, but a peak persists (the effective exponent $\dd \ln \xi_1/ \dd \ln L$ approaches $1$), thus suggesting a disorder-disorder transition (albeit possibly shifting to $T=0$ in the limit $L \rightarrow \infty$).

In summary, while a signature of a first-order transition is observed at $\kappa = 0.48$, a pronounced finite-size dependence of various observables prevents a clear characterization of the range $\kappa^\dagger<\kappa <1/2$. We nevertheless differentiated sizable pre-asymptotic corrections that had previously been (incorrectly) associated with a continuous transition~\cite{kalz2009monte,lee2010study,ramazanov2016thermodynamic}. Particular caution should thus be applied to future studies of this regime.

\subsection{BNNNI model}

For the BNNNI model, we mainly analyze the signatures of the phase transition in the antiphase regime ($\kappa > 1/2$), in order to obtain an overall quantitative phase diagram, which has so far eluded simulation-based approaches.

\subsubsection{Correlation length scaling}
\label{sec:bnnniclength}

\begin{figure}
\includegraphics[width=0.98\columnwidth]{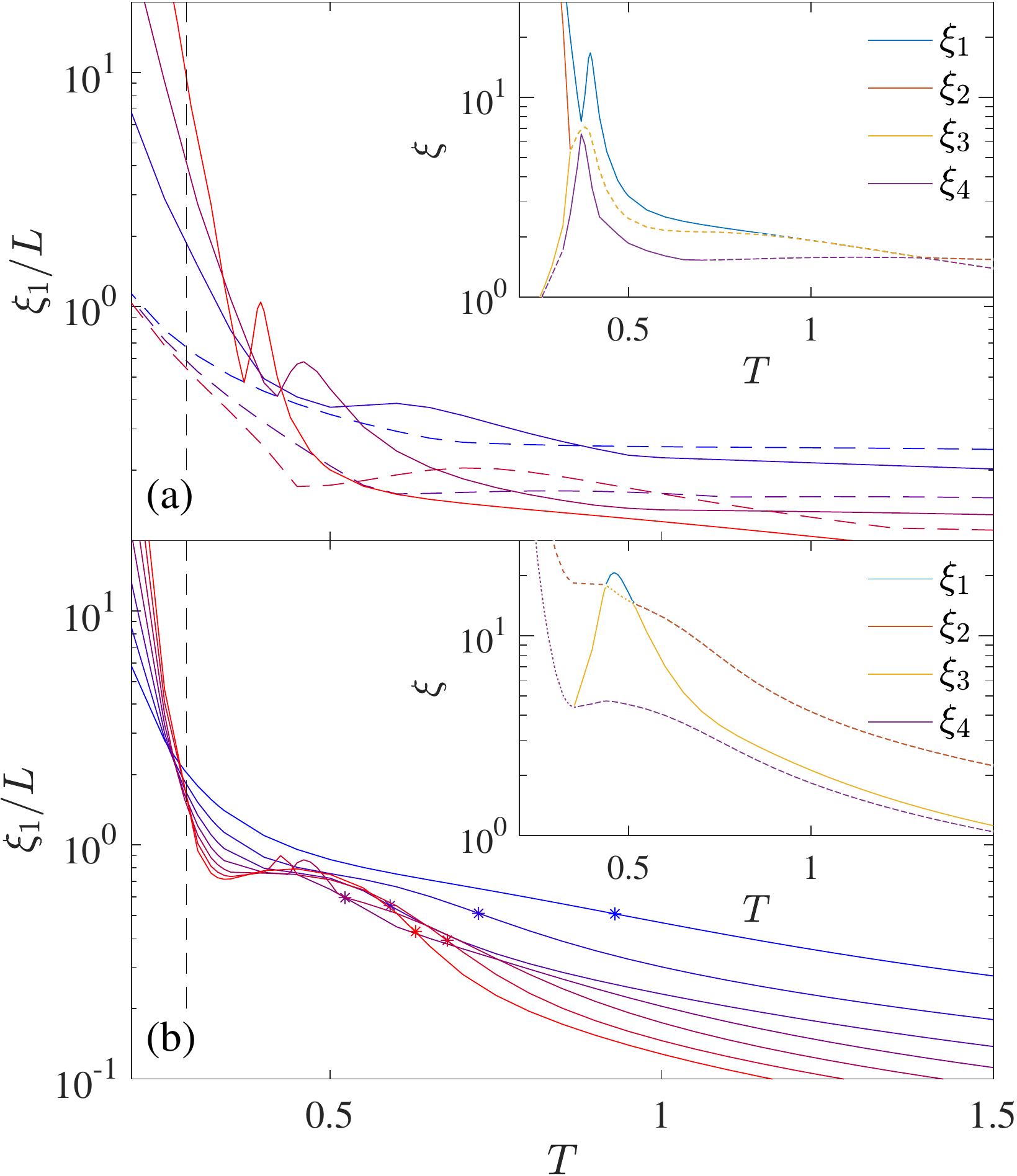}
\caption{Leading correlation length for the BNNNI model with $\kappa=0.6$ in (a) $\paralleltm$ with $L=8, 12, 16$ (solid lines) and $L=6, 10, 14$ (dashed lines); and (b) $\diagtm$ with $L=8, 12, ..., 32$, from blue to red lines, respectively. Asterisks mark the turning point $T^*_1(L)$ used to extrapolate $T_\mathrm{c1}$. Inset: the first four leading correlation lengths in (a) $\paralleltm$ with $L=16$ and (b) $\diagtm$ with $L=24$. Dashed lines denote degenerate correlation length related by complex conjugate subleading eigenvalues. Vertical dashed lines in both panels denote the transition temperature $T_\mathrm{c2} = 0.283(3)$.}
\label{fig:BNNNIclength}
\end{figure}

We first consider the $T$ evolution of correlation lengths with $\paralleltm$ and $\diagtm$ for $\kappa=0.6$ (Fig.~\ref{fig:BNNNIclength}). In both cases, $\xi_1$ evolves non-monotonically as a result of multiple eigenvalue crossings. 
These features are reminiscent of the ANNNI results and markedly differ from their DNNI counterparts, which suggest that a critical phase between $T_\mathrm{c1}$ and $T_\mathrm{c2}$ might be present here as well. 
For $\paralleltm$, a sharp local peak emerges slightly above the antiphase regime for congruent $L~\mathrm{mod}~4 \equiv 0$, and shifts to lower $T$ as $L$ increases. Quantitatively extrapolating transition temperatures from this observable is, however, not realistic given the limited range of accessible system sizes and the size congruence constraint. By contrast, $\diagtm$ presents a much more straightforward trend. As $L \rightarrow \infty$, $\xi_1/L$ (i) diverges in the commensurate antiphase, (ii) approaches a constant in the putative IC phase, $T_\mathrm{c2} \le T \le T_\mathrm{c1}$, and (iii) vanishes in the disordered paramagnetic phase~\cite{nightingale1976scaling,oitmaa1987finite}.
From the lowest temperature crossing points of $\xi_1/L$ between systems of two nearby sizes, $T^*_2(L)$, we can extrapolate $T_\mathrm{c2}$ using a correction form
\begin{equation} \label{eq:BNNNIextrapolateTc}
T^*_2(L) - T_\mathrm{c2} = A_1 L^{-1} ( 1 + A_2 L^{-1} ),
\end{equation}
where $A_1$ and $A_2$ are fitted constants. Although the scaling of the correction is not known a priori, the resulting extrapolation is nearly linear for all $\kappa$ between $0.55$ and $2$, thus giving credence to this form. For $\kappa=0.6$, in particular, fitting results from $L=14$ to $30$ gives $T_\mathrm{c2} = 0.283(3)$. 

While for the ANNNI model $T_\mathrm{c1}$ can be determined from the finite-size scaling of the local exponent $Y_L = \delta \ln \xi_1 / \delta \ln L$ via the $\paralleltm$ approach~\cite{hu2021resolving},
here the situation is not as straightforward. Eigenvalue crossings indeed persist even in the $\diagtm$ analysis [Fig.~\ref{fig:ANNNIclength}(b)].
As a result, a bump-like peak appears in the plateau regime for $L= 24, 28$ but retreats for $L=32$. (For $L=32$, a second peak in $\xi_3$ appears but does not cross $\xi_1$ nor $\xi_2$.) In order to understand the physical origin of this effect, recall that the angular argument of this pair of complex conjugate subleading eigenvalues corresponds to the modulation period. A sharp peak thus corresponds to the presence of a real eigenvalue (with wavenumber $q=\mathrm{arg}(\lambda_1)/(2\pi) = 0$).
Because the modulation can propagate along either diagonal directions, this peak indicates that the modulation perpendicular to the TM propagation direction momentarily dominates within the relevant temperature window. Said differently, $L$ is then commensurate with the preferred wavenumber. The overall scaling trend, however, remains unaffected. 

The numerical challenge of determining $T_{c1}$ nevertheless remains. A first option is to consider the scaling of anisotropy, as for the ANNNI model~\cite{beale1985finite,hu2021resolving}. Although small systems display a smooth evolution~\cite{oitmaa1987finite}, larger $L$ exhibit complex oscillations. The correlation length scaling in the IC phase regime is thus severely affected by the choice of boundary condition---again possibly resulting from the interference between preferred wavenumber and $L$---hence preventing a clear determination of $T_\mathrm{c1}$.
A second option is to restrict modulation to lie along the TM direction of propagation by examining the leading correlation length associated with a complex eigenvalue, $\xi'_1$.
Given the smooth evolution of $\xi'_1$ with $L$, it is then possible to extrapolate $T_\mathrm{c1}$. As a finite-$L$ echo of $T_\mathrm{c1}$, we consider the local minimum 
\begin{equation}
\left.\frac{\partial(\ln \xi'_1)}{\partial T}\right|_{T^*_1(L)}=0, 
\end{equation}
which graphically corresponds to the turning point of $T$-$\ln(\xi'_1/L)$. As $L$ increases, $T^*_1(L)$ first decreases but the monotonic trend does not persist (e.g., $T^*_1(L=28) > T^*_1(L=24)$). As $\kappa$ increases, both the $\xi_1$ plateau and its turning onset weaken. Because eigenvalue crossing is absent for the largest $L$ considered and $T^*_1(L)$ evolves smoothly and monotonically, a tentative extrapolation of $T_\mathrm{c1}$ using Eq.~\eqref{eq:BNNNIextrapolateTc} is possible for $\kappa \ge 0.8$ (Fig.~\ref{fig:BNNNIdiag}). We thus have that $T_\mathrm{c1}>T_\mathrm{c2}$, beyond the uncertainty range, up to $\kappa=1.5$, and the signature of the critical phase is qualitatively visible up to $\kappa = 2$. This evidence marginally supports the absence of a Lifshitz point, i.e., $\kappa^*=\infty$. 
The ANNNI and the BNNNI models both exhibit a similar correlation length plateau which suggests that an IC phase exists in the former as well~\cite{hu2021resolving}. We should note, however, our $T_\mathrm{c1}$ values are tentative because similar finite-size corrections---non-monotonic trend of $T^*_1(L)$---may be observed at larger $L$ (although less severely). Hence we also cannot exclude as strongly as for the ANNNI model that the BNNNI IC-like phase disappears in the thermodynamic limit.

\subsubsection{Heat capacity evolution} 

\begin{figure}
\includegraphics[width=0.98\columnwidth]{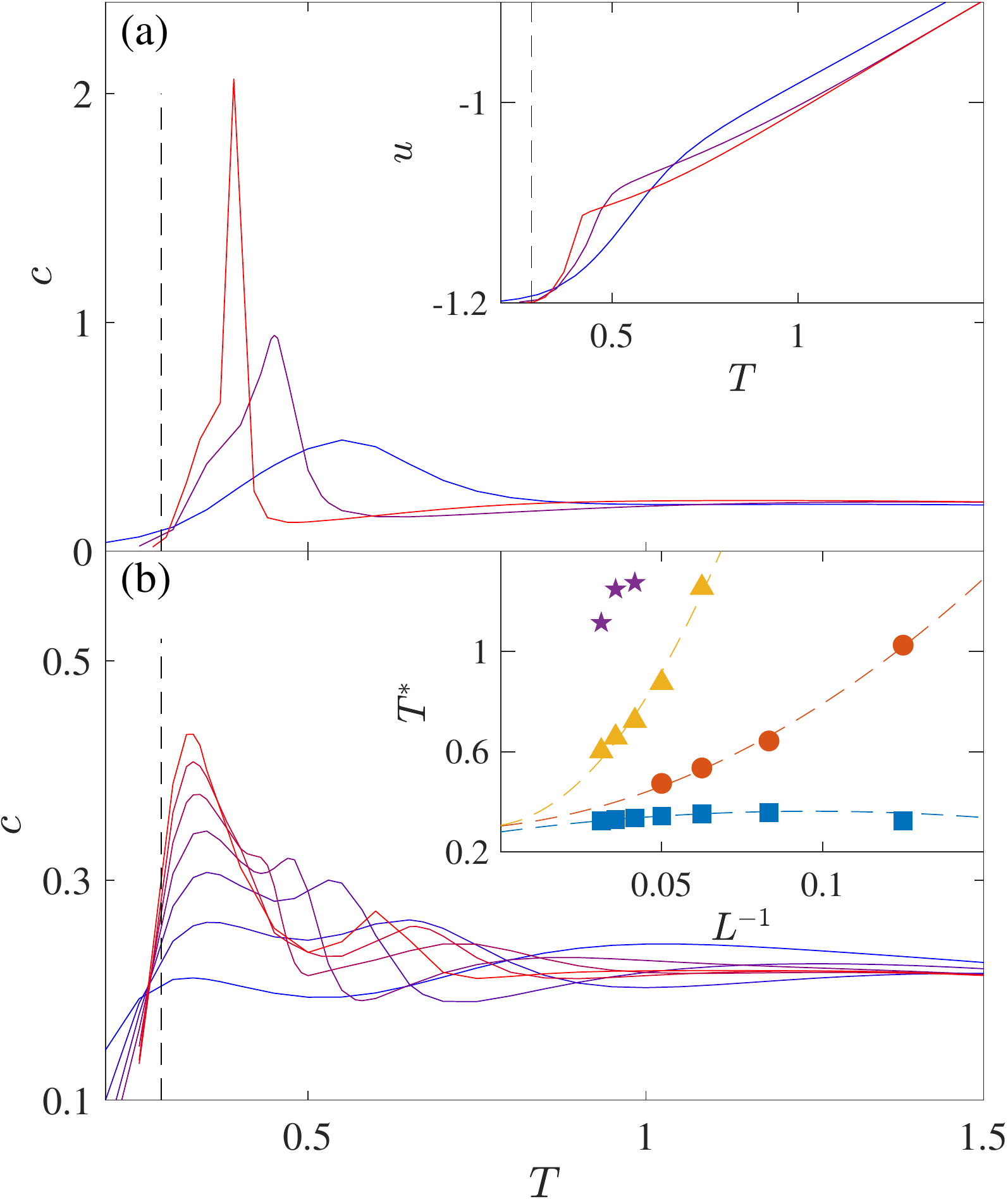}
\caption{Specific heat for the BNNNI model for $\kappa=0.6$ in (a) $\paralleltm$ for $L=8, 12, 16$ and (b) $\diagtm$ for $L=8, 12, ..., 32$, from blue to red. Vertical dashed lines denote $T_\mathrm{c2} = 0.283(3)$. For (a), the sharp peak is located at marginally higher $T$ than the local peak of $\xi_1$ [Fig.~\ref{fig:BNNNIclength}(a)] by $\sim -1\%$ differences in $L=12,16$. Insets of (a): energy per spin for the same systems; inset of (b): specific heat peak temperatures are extrapolated (using a quadratic fitting form) to merge at $T_\mathrm{c2} \approx 0.3$ as $1/L \rightarrow 0$.}
\label{fig:BNNNIthermo}
\end{figure}

We next investigate the evolution of the heat capacity. Because $\paralleltm$ and $\diagtm$ exhibit different finite-size features, we consider both.
For $\paralleltm$, a sharp peak grows with $L$, but its temperature is marginally smaller than that of the local peak of $\xi_1$. Correspondingly, the internal energy grows stepwise with a decreasing step height as $L$ increases, as observed in Monte Carlo simulations~\cite{velgakis1988critical} and in the $\perptm$ solution of the ANNNI model~\cite{hu2021resolving}. 
This behavior is consistent with a Pokrovsky-Talapov type transition, at which the heat capacity divergence is discontinuous (with scaling exponent $\alpha' \rightarrow \infty$) from the antiphase side.
For $\diagtm$, the $c(T)$ curve is multiply peaked. The lowest temperature peak is the highest, whereas higher temperature peaks grow and shift to lower $T$ as $L$ increases. These peaks appear to evolve toward $T_\mathrm{c2}$ [Inset of Fig.~\ref{fig:BNNNIthermo}(b)], as they do as finite-$L$-echo of the IC phase in the ANNNI model~\cite{hu2021resolving}. 
The analogy between the two models suggests that for $\kappa > 1/2$ the antiphase of the BNNNI model also undergoes a Pokrovsky-Talapov transition~\cite{pokrovsky1979ground} at $T_\mathrm{c2}$, followed by critical IC phase~\cite{kosterlitz1973ordering} that terminates at a KT transition at $T_\mathrm{c1}$. The IC phase, if it exists, would then be characterized by an algebraically diverging correlation length and presents a stepwise evolution of the modulation in finite systems.

\subsubsection{Phase diagram}

\begin{figure}[ht]
\includegraphics[width=0.98\columnwidth]{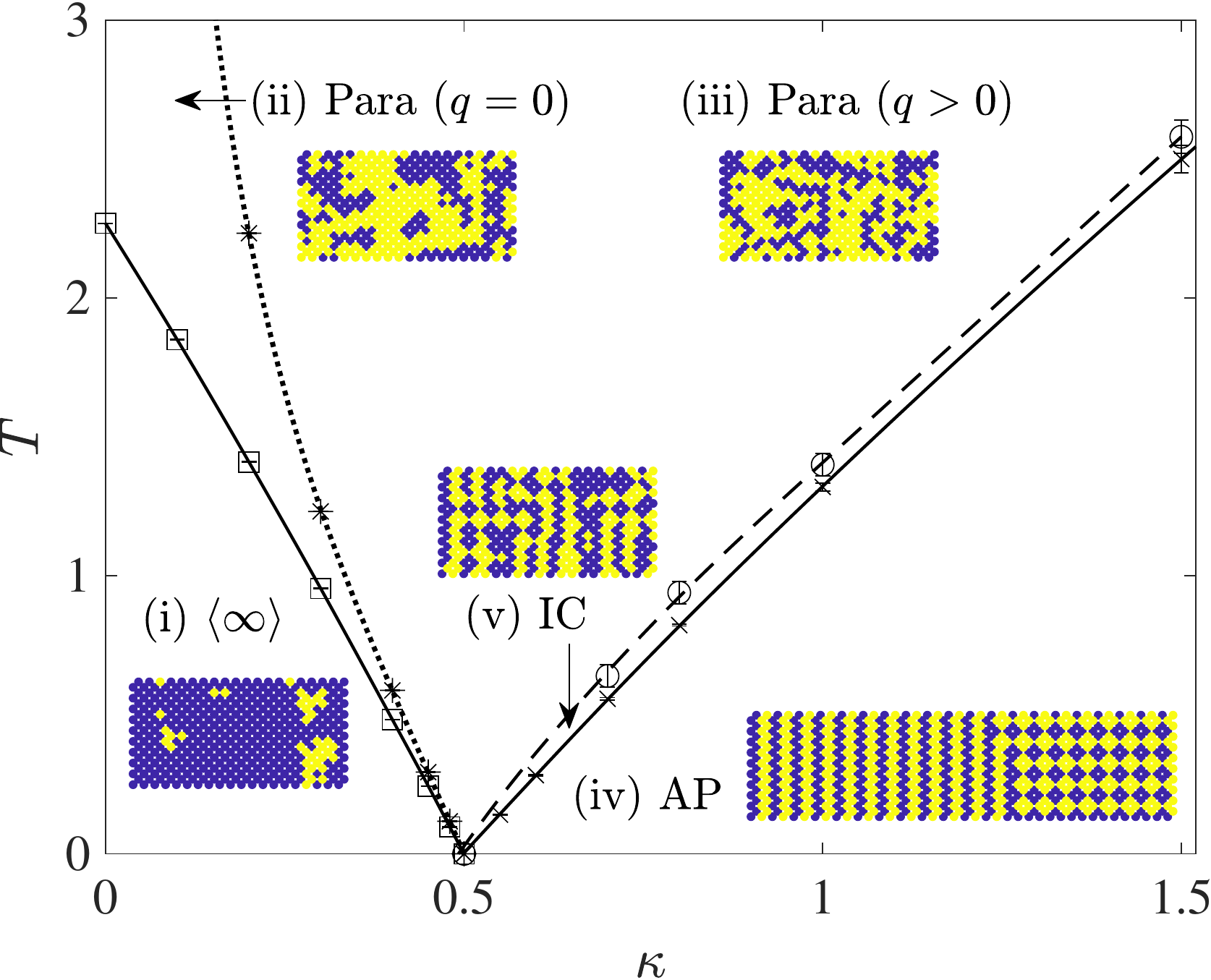}
\caption{Phase diagram for the BNNNI model. The TM approach provides phase boundaries for the ferromagnetic $\langle \infty \rangle$-paramagnetic (squares), degenerate antiphases (AP) to IC (crosses), and paramagnetic-IC (circles) transition. The disorder line (asterisks) subdivides the paramagnetic phase into $q=0$ and $q>0$ modulation wavenumbers. Configuration snapshots generated by planting~\cite{hu2020comment} with $\diagtm$ use blue and yellow points to denote $+1$ and $-1$ spins, respectively.}
\label{fig:BNNNIdiag}
\end{figure}

Combining the correlation length and heat capacity results offers a consistent phase diagram of the BNNNI model (see Fig.~\ref{fig:BNNNIdiag}). The simple ferromagnetic regime at $\kappa<1/2$ presents an Ising-type transition at $T_\mathrm{c}$, as identified by $\xi_1/L$ crossings, to the paramagnetic phase. (Quantitative estimates are fully consistent with earlier TM~\cite{oitmaa1987finite} and free fermion approximation~\cite{dasgupta1991bnnni} results.) An additional disorder line can be identified from the splitting of subleading eigenvalues from a pair of complex conjugates (at high $T$) into two distinct real numbers (at low $T$). Being non-critical, these two lines are only marginally affected by finite-size corrections~\cite{oitmaa1987finite}. For $\kappa>1/2$, two transitions can be identified, $T_\mathrm{c2}<T_\mathrm{c1}$, as discussed in Sec.~\ref{sec:bnnniclength}. Prior estimates for $T_\mathrm{c2}$ vary dramatically, but our results robustly fall between those of Ref.~\onlinecite{landau1985phase} and those of Ref.~\onlinecite{dasgupta1991bnnni}. Around $\kappa = 1/2$, the TM approach suggests that the phase boundary has a finite slope on both sides of the multicritical point, thus supporting the free fermion approximation results over those of the renormalization group approach~\cite{aydin1989renormalisation}. 
For $T_\mathrm{c1}$, various qualitative proposals have been made~\cite{aydin1989monte,aydin1989renormalisation}, but to the best of our knowledge no quantitative estimates were reported. Our results, albeit still somewhat imprecise, are consistent with the two-step melting scenario persisting over a wide range of $\kappa$ and the presence of intermediate critical IC phase.

\subsection{3NNI model}
\label{sec:result:3nn}

\begin{figure}[t]
\includegraphics[width=0.98\columnwidth]{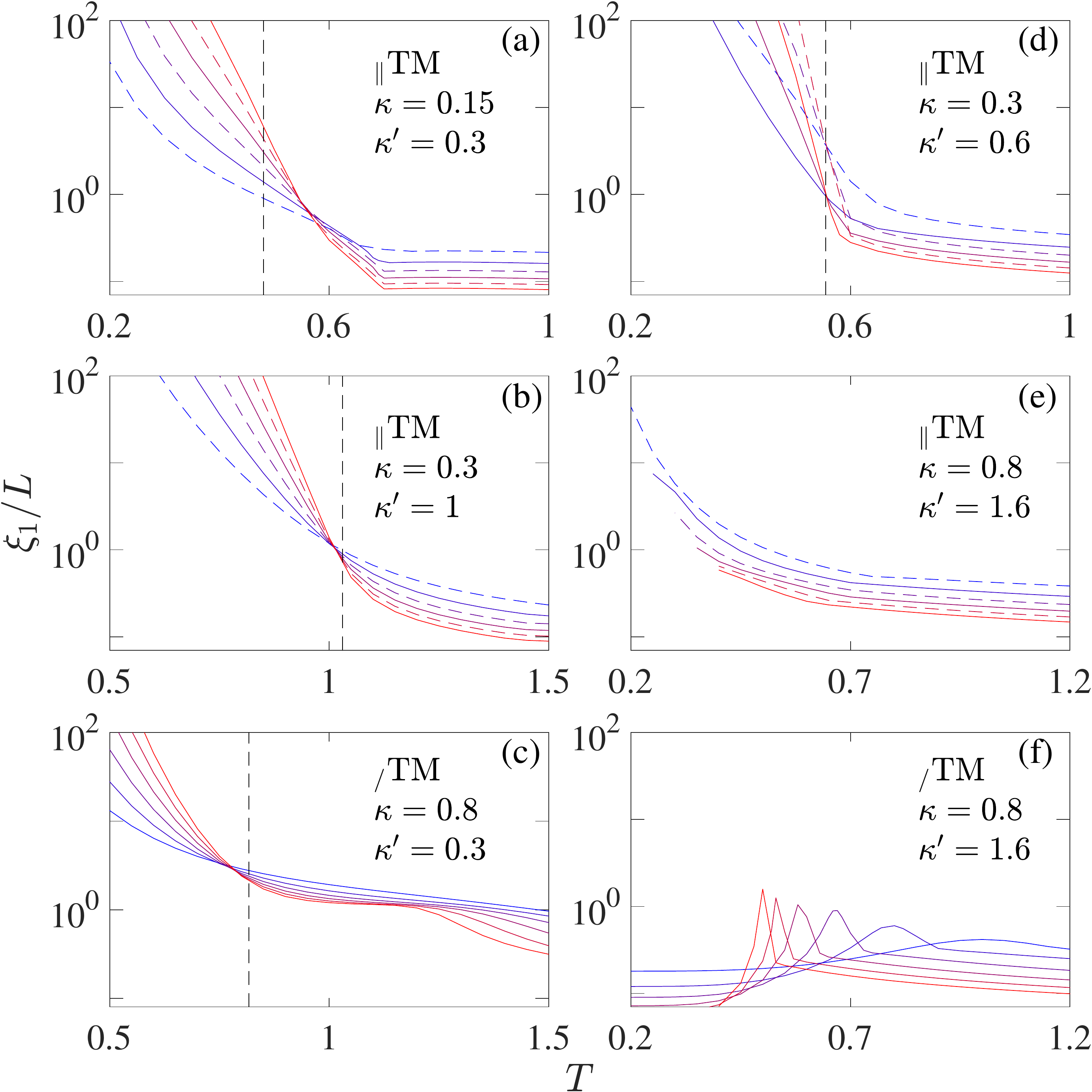}
\caption{Correlation length scaling of the 3NNI model for different frustration parameters. For $\paralleltm$, solid and dashed lines denote systems with $L=8,12,16$ and $L=6,10,14$, from blue to red, respectively. For $\diagtm$, lines denote $L=8,12,...,28$, from blue to red. In (a-d), the ground states are ferromagnetic, $\langle 1\rangle$, $4 \times 4 $ and $\langle 2 \rangle$, respectively [See Fig.~\ref{fig:intro-model}(d)]. In (e, f), the ground state is disordered. Note that results in (e) are truncated at large $L$ and low $T$, because numerical instability of the eigensolver then gives rise to complex $\lambda_0$.}
\label{fig:3NNclength}
\end{figure}

\begin{figure}[ht]
\includegraphics[width=0.98\columnwidth]{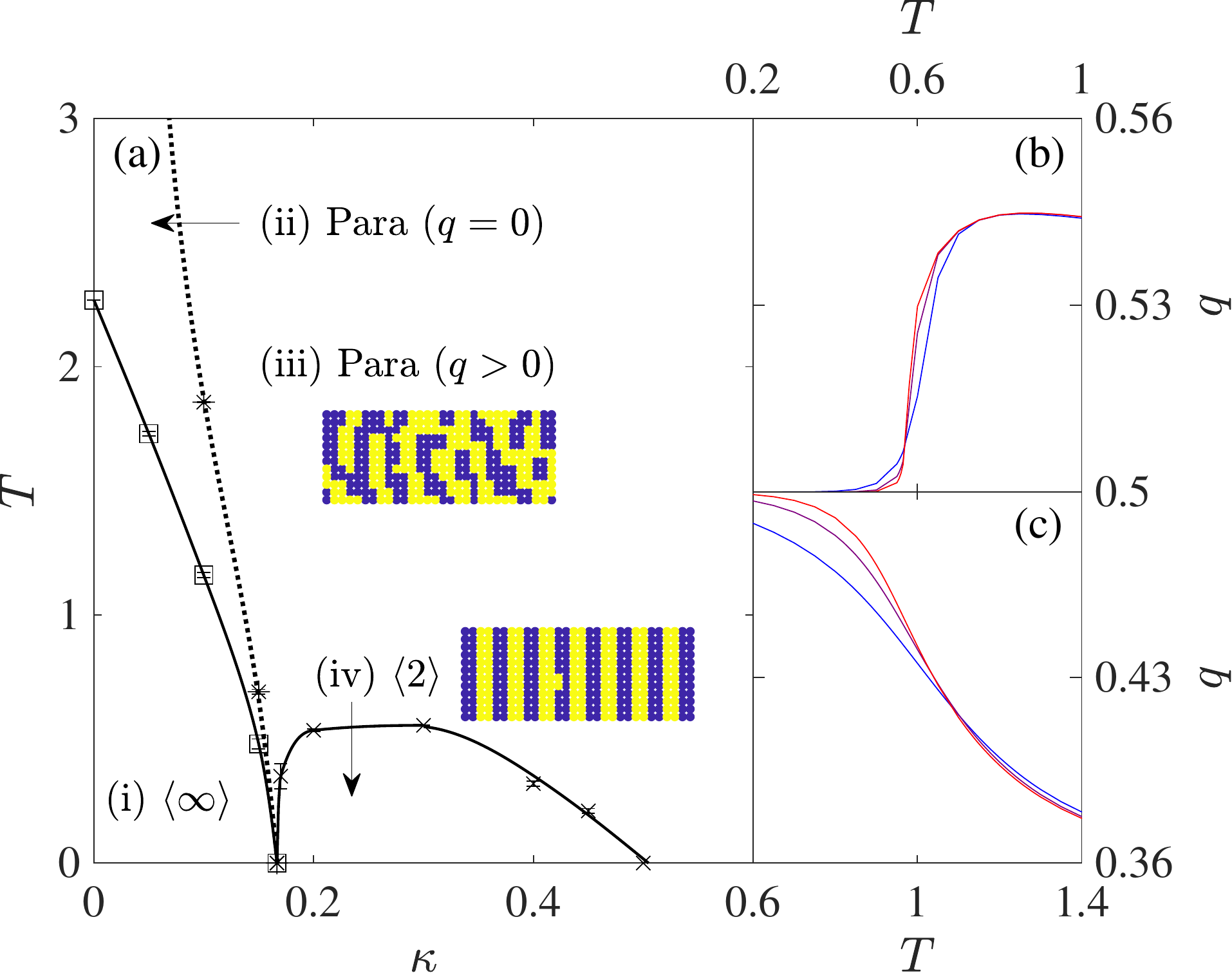}
\caption{(a) Phase diagram for the 3NNI model for $\kappa=\kappa'/2$. The TM approach provides phase boundaries for the ferromagnetic $\langle \infty \rangle$-paramagnetic (squares), $\langle 2 \rangle$ antiphases to paramagnetic (crosses), and the disorder line (asterisks) subdivides the paramagnetic phase into $q=0$ and $q>0$ modulation wavenumbers. Configuration snapshots generated by planting~\cite{hu2020comment} with $\paralleltm$ use blue and yellow dots to denote $+1$ and $-1$ spins, respectively. (b, c) The modulation wavenumber $q$ given by $\paralleltm$ for (b) the 3NNI model with $\kappa=0.15, \kappa'=0.3$, and (c) ANNNI model with $\kappa=0.6$ in $L=8,12,16$, from blue to red.}
\label{fig:3NNdiagA}
\end{figure}

We finally consider the generic 3NNI model. As shown in Fig.~\ref{fig:intro-model}(d), the ground state configuration of this model depends on both $\kappa$ and $\kappa'$. The model is thus expected to present different types of order-disorder transitions, as characterized by the correlation length scaling (Fig.~\ref{fig:3NNclength}). In parameter regimes corresponding to the ferromagnetic, $\langle 1 \rangle$ or $4 \times 4 $ ground state configurations, these lengths indeed behave distinctly. Different regimes analogous to those observed in other models can further be identified (Fig.~\ref{fig:3NNclength}). 
\begin{itemize}
\item In the ferromagnetic regime, $\xi_1/L$ curves cross then kink as $T$ increases, as they would for the thermodynamic phase transition $T_\mathrm{c}$ and for the disorder line crossover of the BNNNI model (the dotted line in Fig.~\ref{fig:BNNNIdiag}), respectively.
\item In the $\langle 1 \rangle$ regime, $\xi_1/L$ decays monotonically and presents a single fixed point. The order of phase transition as well as the critical exponent values in the continuous transition regime also vary with the choice of $(\kappa, \kappa')$~\cite{liu2016role}, as in the DNNI model [Fig.~\ref{fig:DNNIthermo}(a)].
\item In the $4 \times 4 $ regime, the $\xi_1/L$ curves from $\diagtm$ plateau after the crossing, as in the BNNNI model [Fig.~\ref{fig:BNNNIclength}(b)].
\end{itemize}

The $\langle 2 \rangle$ phase of the 3NNI model, however, melts differently from that of the ANNNI model. As shown in [Fig.~\ref{fig:3NNclength}(d)], $\xi_1/L$ curves cross at single point (for congruent $L \mod 4$), then decay monotonically without exhibiting any shoulder [cf. Fig.~\ref{fig:ANNNIclength}(b)].
This transition has been identified as being first-order~\cite{landau1985phase,liu2016role}, but the distinction between a first-order and a Pokrovsky-Talapov transition is ambiguous in Monte Carlo simulations~\cite{landau1985phase,velgakis1988critical}. These two features here clearly support the former over the latter.

As mentioned in Sec.~\ref{sec:model:3NN}, the three-dimensional version of the 3NNI model was proposed as a minimal model of microemulsions. Frustration parameters were then set to $\kappa = \kappa'/2$ in order to match the oil-water-surfactant representation~\cite{widom1986lattice,dawson1988phase}. Interestingly, in two dimension this parameter choice results in crossing $\langle 2 \rangle$---lamellar-like---regime, and then following the $\langle 1 \rangle$--$4 \times 4$ boundary, at which the system is always disordered, as frustration increases.
We thus here consider the large frustration regime ($\kappa=\kappa'/2 >1/2$).
For $\paralleltm$ [Fig.~\ref{fig:3NNclength}(e)], $\xi_1/L$ grows monotonically with decreasing $T$ but no crossing is detected. Presumably $\xi_1/L$ then diverges at a zero temperature phase transition.
For $\diagtm$ [Fig.~\ref{fig:3NNclength}(f)], however, the $\xi_1/L$ peak shifts to lower $T$ at large $L$, a behavior reminiscent of what happens at $\kappa=1/2$ in the DNNI model [Fig.~\ref{fig:DNNIthermoq}(c)]. 
Moreover, in the limit $\kappa (= \kappa'/2) \rightarrow \infty$, the model reduces to two penetrating and decoupled DNNI antiferromagnets with $\kappa=1/2$. The $\paralleltm$ and $\diagtm$ approaches for the 3NNI model are then equivalent to the $\diagtm$ and $\perptm$ approaches for the DNNI model, respectively. The model is thus always disordered beyond a zero-temperature phase transition~\cite{landau1980phase}.

Figure~\ref{fig:3NNdiagA}(a) presents a sketch of the 3NNI $\kappa$-$T$ phase diagram at $\kappa=\kappa'/2$.
The emergence of lamellar microphases at intermediate frustration is characteristics of SALR microphase formers~\cite{almarza2014periodic,zhuang2016equilibriumPRL,zhuang2016recent,hu2018clustering}. 
Specifically, a single first-order transition bounds the $\langle 2 \rangle$ phase between $1/6 < \kappa < 1/2$. The wavenumber $q$ along the axial direction jumps at the transition [Fig.~\ref{fig:3NNdiagA}(b)], as expected of a first-order transition scenario. This behavior sharply contrasts with the stepwise decrease of $q$ in the ANNNI model [Fig.~\ref{fig:3NNdiagA}(c)], which is expected to follow a square-root singularity in the thermodynamic limit~\cite{selke1988annni,sato1999equilibrium,derian2006modulation}. 
The disordered phases in these two models are also morphologically different. In the 3NNI model, spin clusters of width 2 ($\uparrow \uparrow$) echo the dissolved modulation. For the ANNNI model, spins instead form layers of width $>2$ in the vicinity of $T > T_\mathrm{c1}$~\cite{hu2021resolving}, echoing the floating IC phase. In summary, the first-order scenario for the 3NNI model at $\kappa=\kappa'/2$ is reminiscent of its three-dimensional counterpart, which also exhibit a weakly first-order transition at the melting of the modulated phase~\cite{dawson1988phase}. 

\section{Conclusions} \label{sec:conclusion}

Using a numerical TM approach, we have resolved various long-standing questions about the phase behavior of a series of two-dimensional frustrated Ising models. 
For the ANNNI model, our consideration of the domain-wall free energy supports the existence of the critical IC phase, thus extending our recent analysis~\cite{hu2021resolving}. 
For the DNNI model, the TM results confirm the location of the transition in the limiting case $T_\mathrm{c}(\kappa=1/2)=0$, and support and distinguish the weakly first-order transition scenario for $\kappa^\dagger < \kappa <1/2$ and for $1/2<\kappa<\kappa^*$. 
For the BNNNI model, a strong signature of the critical IC phase is identified, even though its upper boundary remains imprecise. 
For the 3NNI model, the lamellar modulated regime has been shown to melt with a single first-order transition, in contrast to that of the ANNNI model.
Combining these findings provides a systematic overview of modulated phase formation, and high-accuracy benchmarks for other theoretical and numerical approaches. 

The numerical TM method nevertheless still suffers from an insufficiently wide range of system sizes under certain circumstances, such as determining $T_\mathrm{c1}$ for the BNNNI model. Some of these problems might be resolved, in time, thanks to ever improving computers architecture. Relaxing exactness, such as by using inexact eigensolvers~\cite{wang2020efficient} or truncated configuration representations as in the DMRG approach~\cite{nishino1995density,derian2006modulation} might be more time effective. More immediately, the TM approach could certainly be used to other lattices models such as spin-$1$ and Potts models. 

\begin{acknowledgements}
We acknowledge support from the National Science Foundation Grant No. DMR-1749374 and from the Simons Foundation (\#454937).
The computations were carried out on the Duke Compute Cluster and on Extreme Science and Engineering Discovery Environment (XSEDE), which is supported by National Science Foundation grant number ACI-1548562.
Data relevant to this work have been archived and can be accessed at the Duke Digital Repository~\cite{lpdata}.
\end{acknowledgements}

%%%%%%%%%%%%%%%%%%%%%%%%%%%%%%%%%%%%%%%%%%%%%%%%%%%%
%
% Appendix begins
%
%%%%%%%%%%%%%%%%%%%%%%%%%%%%%%%%%%%%%%%%%%%%%%%%%%%%
\appendix

\section{Decomposition for the matrix-vector multiplication} \label{appd:construction}

In this Appendix we detail the transfer matrix decomposition scheme used for two-dimensional frustrated Ising models. Although this approach was first implemented in the 1980s~\cite{pesch1985transfer,beale1985finite,oitmaa1987finite}, earlier reports omitted most technical details. This Appendix and the following aim to fill this gap, and thus facilitate future extensions of these methods.

The general strategy is as follows. Although transfer matrix entries are straightforwardly expressed (Eq.~\eqref{eq:tmatgeneral}), storing the whole matrix $\mathbf{T}$ in memory becomes quickly beyond practical reach as $L$ increases. The solution relies on using iterative eigenvalue algorithms (such as power iteration or Krylov subspace-based iterations) that only require a matrix-vector multiplication subroutine, with vector $\mathbf{v}$ as input and $\mathbf{w} = \mathbf{T}\mathbf{v}$ as output, and thus avoid explicitly storing $\mathbf{T}$. Because $\mathbf{T}$ is structured, we further decompose it into a product of sparse matrices, which saves additional memory space as well as computer time. In this Appendix, we first introduce the algorithm in the context of ANNNI model in $d=2$, and then generalize it to the DNNI and 3NNI (include BNNNI) models.

\begin{figure*}
\includegraphics[width=0.86\textwidth]{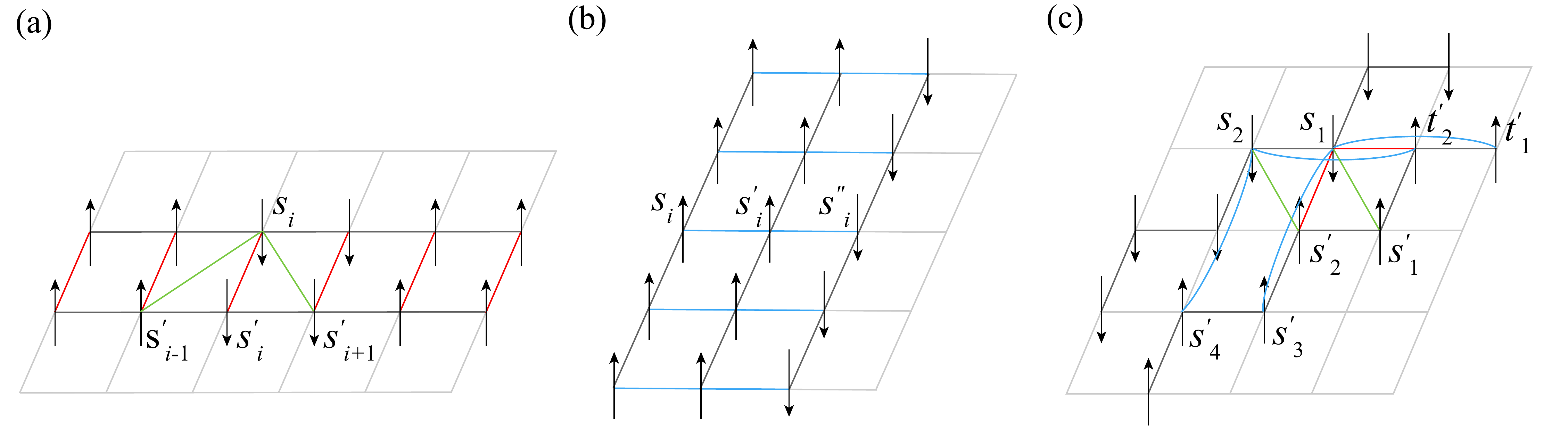}
\caption{Schematics of inter-layer interaction in (a) $\tmat{\perp,\mathrm{A}}{\mathrm{TM}}$ and $\tmat{\perp,\mathrm{D}}{\mathrm{TM}}$, (b) $\tmat{\parallel,\mathrm{A}}{\mathrm{TM}}$ and $\tmat{\parallel,\mathrm{3}}{\mathrm{TM}}$, (c) $\tmat{/,3}{\mathrm{TM}}$. Black lines denote spin layers; red, blue and green lines denote the inter-layer Ising nearest-neighbor interaction, (bi)axial NNN interaction, and diagonal interactions, respectively. Note that for (c) the inter-layer Ising nearest-neighbor interaction for $(\mathbf{s}, \mathbf{s'})$ involves every other spin $s_2, s_4, ...$ on a layer.}
\label{fig:tmatpropagation}
\end{figure*}

\subsection{ANNNI model propagated perpendicular to the axial direction} \label{sec:2XAT}
We first consider the $\perptm$ case for the ANNNI model [Fig.~\ref{fig:tmatpropagation}(a)]. In this case, we denote the spin layer state $\mathbf{s} = (\{\pm 1\}^L)$ as an $L$ dimensional binary vector, which is encoded with an $L$-bit unsigned integer, $a$, so that $0 \le a \le 2^L-1$ naturally include all possible layer configurations with $a_1, ..., a_L \in \{0, 1\}$. The physical state $\mathbf{s}$ and machine-expressed state $a$ is related by the simple mapping $a_i = 1 \rightarrow s_i = 1$ and $a_i = 0 \rightarrow s_i = -1$. 

The energetic contribution of intra-layer interactions of this state is then
\begin{equation}
V_{x} (a) = -J \sum_{i=1}^{L} s_i s_{i+1} + \kappa J \sum_{i=1}^{L} s_i s_{i+2} - h \sum_{i=1}^{L} s_i.
\end{equation} 
Technically, this expression can be evaluated using bitwise operations. For convenience we first define the net number of positive spins as a function of $a$, such that $\netp(a) = 2\popc(a) - L$, where $\popc$ counts the bits set to 1. One can then write
\begin{equation} \begin{aligned}
V_x (a) = J \netp(a \bitxor \rol(a, 1)) &- \kappa J \netp(a \bitxor \rol(a, 2)) \\ &- h \netp(a),
\end{aligned} \end{equation}
where $\rol$ is rotate-left-shift ($\ror$ is similarly rotate-right-shift) and $\bitxor$ is bitwise xor. Similarly, the contribution of the energy of the neighboring layer reads
\begin{equation} \begin{aligned}
V_z(a, a') &= -J \sum_{i=1}^{L} s_i s'_i \\
&= J \netp(a \bitxor a').
\end{aligned} \end{equation}

Formally, the transfer matrix $\mathbf{\mathring{T}}$ of size $N_\mathrm{states}=2^L$ (shorthand for $\tmat{{\perp},A}{\mathring{T}}$) has entries
\begin{equation}
\mathbf{\mathring{T}}_{aa'} = e^{-\beta (V_x(a) + V_z (a, a'))} = \mathbf{T}^x_{a,a} \times \mathbf{T}^z_{a,a'}.
\end{equation}
The binary representation of the entry index $a, a'$ naturally gives the spin configuration of the corresponding layer state, as described above. $\mathbf{\mathring{T}}$ can further be decomposed into a diagonal matrix $\mathbf{T}^x$ with entries $\mathbf{T}^x_{a,a} = e^{-\beta V_x(a)}$ and a symmetric (and centrosymmetric) matrix $\mathbf{T}^z$ with entries $\mathbf{T}^z_{a,a'} = e^{-\beta V_z(a, a')}$. 

The partition function of an $N$-layer system can be expressed using $\tr(\mathbf{\mathring{T}}^N)$, which however is not symmetric. To leverage the efficiency of fast numerical eigensolvers for symmetric matrices, we define an alternate symmetric transfer matrix
\begin{equation} \label{eq:tmatdecomp}
\mathbf{T} = (\mathbf{T}^x)^{1/2} \mathbf{T}^z (\mathbf{T}^x)^{1/2}.
\end{equation}
which has the same the eigenvalues as the original matrix, and eigenvectors related by
\begin{equation}
\begin{cases}
\varphi(\mathbf{T}) &= (\mathbf{T}^x)^{-1/2} \varphi(\mathbf{\mathring{T}}), \\
\varphi^{-1}(\mathbf{T}) &= \varphi^{-1}(\mathbf{\mathring{T}}) (\mathbf{T}^x)^{1/2}, \\
\end{cases}
\end{equation}
where the superscript $-1$ denotes the left eigenvector. 
In zero external field, $h=0$, $\mathbf{T}$ is further centrosymmetric ($\mathbf{T}_{a,a'} = \mathbf{T}_{2^L-a-1,2^L-a'-1}$). As a result of this transformation, we have that: (i) all eigenvalues are real; (ii) all eigenvectors are orthogonal; (iii) every eigenvector is either symmetric or skew-symmetric~\cite{cantoni1976eigenvalues} when $h=0$.

Because the size of $\mathbf{T}$ grows exponentially with $L$, storing the full matrix in memory becomes first inefficient and then impractical as $L$ increases. However, a subroutine that computes matrix-vector multiplications on the fly can be used to extract the first several leading eigenvalues and eigenvectors. A direct multiplication requires $\OO(4^L)$ arithmetic operations, but can be reduced by factorizing $\mathbf{T}^z$ into sparse matrices as~\cite{blote1982critical} 
\begin{equation} \label{eq:Tzfact}
\mathbf{T}^z = \mathbf{T}^{z,L} \mathbf{T}^{z, L-1} \cdots \mathbf{T}^{z,1},
\end{equation}
where $\mathbf{T}^{z,i}$ has two nonzero entries in each row,
\begin{equation} \label{eq:Tz1def}
\begin{cases}
\mathbf{T}^{z,i}_{a,a} = e^{\beta J}, \\
\mathbf{T}^{z,i}_{a, a'} = e^{-\beta J}.
\end{cases}
\end{equation}
The off-diagonal indexes $a'=a \bitxor \rol(1, i-1)$ here denote the configurations obtained by flipping the $i$-th spin from $a$. Note that $\mathbf{T}^{x,i}$ is transformed from $\mathbf{T}^{z,1}$ by re-indexing $a$ to $\rol(a, i-1)$, formulated by the permutation operation 
\begin{equation} \label{eq:Tzpermut}
\mathbf{T}^{z,i} = (\mathbf{P}^T)^{i-1} \mathbf{T}^{z,1} \mathbf{P}^{i-1} 
\end{equation}
where $\mathbf{P}^i$ is the permutation matrix with indexes $(a, a')$ to be $1$ and 0 otherwise. Inserting Eq.~\eqref{eq:Tzpermut} into Eq.~\eqref{eq:Tzfact} and knowing $\mathbf{P}^T \mathbf{P} = \mathbf{I}; \mathbf{P}^L = \mathbf{I}$ gives
\begin{equation} \label{eq:xa2tzsimplify}
\mathbf{T}^{z} = (\mathbf{P} \mathbf{T}^{z,1} )^L.
\end{equation} 
In summary, the matrix-vector multiplication can be conducted by a sequence of multiplication with sparse matrices
\begin{equation} \label{eq:matvecannni2d}
\mathbf{T} \mathbf{v} = (\mathbf{T}^x)^{1/2} (\mathbf{P} \mathbf{T}^{z,1} )^L (\mathbf{T}^x)^{1/2} \mathbf{v}
\end{equation}
with complexity $\OO( 2^L L)$.

In addition, we have $\mathbf{T}$ is invariant under the permutation of circularly shifting one spin, such that
\begin{equation}
\mathbf{T} = \mathbf{P}^T \mathbf{T} \mathbf{P},
\end{equation}
where $\mathbf{P}^L = \mathbf{I}$. If $\varphi$ is an eigenvector of $\mathbf{T}$, then $\mathbf{P} \varphi$ is also an eigenvector of $\mathbf{T}$ associated with same eigenvalue $\lambda$, because
\begin{equation}
\mathbf{T} (\mathbf{P} \varphi) = \mathbf{P} \mathbf{T} \varphi = \lambda \mathbf{P} \varphi.
\end{equation}
As a result, an eigenvector of non-degenerate eigenvalue is invariant under the permutation of $\mathbf{P}$, and degenerate eigenvectors associated with degenerate eigenvalue (and their linear combinations) form a cyclic group. This structural property can be used to optimize the extraction of leading eigenvalues, as described in Appendix~\ref{appd:reducedtmat}.

\subsection{ANNNI model propagated along the axial direction} \label{sec:2YAT}

We now consider $\paralleltm$ for the ANNNI model [Fig.~\ref{fig:tmatpropagation}(b)]. In this case, we denote three subsequent layers as $a,a',a''$ where the state of each layer is encoded as an $L$-bit integer, as above. The energetic contribution of $a$ then includes NN interactions with itself and with $a'$ as well as NNN interaction with $a''$,
\begin{equation}
\begin{cases}
V_{x} (a,a') &= -J \sum_{i=1}^{L} s_i s_{i+1} - J \sum_{i=1}^{L} s_i s'_i \\ &\quad - h \sum_{i=1}^{L} s_i,\\
V_{z} (a,a'') &= \kappa J \sum_{i=1}^{L} s_i s''_i.\\
\end{cases}
\end{equation}
The bitwise operations then reads
\begin{equation} \label{eq:yannnipotentials}
\begin{cases}
V_x (a,a') &= J \netp(a \bitxor \rol(a, 1)) + J \netp(a \bitxor a') \\ &\quad - h \netp(a),\\
V_z (a,a') &= -\kappa J \netp(a \bitxor a'').
\end{cases}
\end{equation}

We then define the transfer matrix $\mathbf{T}$ (shorthand for $\tmat{\parallel,A}{T}$) with entries
\begin{equation} \begin{aligned}
\mathbf{T}_{(a,a'),(a',a'')} &= e^{-\beta (V_x(a,a') + V_z (a, a''))} \\ &= \mathbf{T}^x_{(a,a'),(a,a')} \cdot \mathbf{T}^z_{(a,a'),(a',a'')},
\end{aligned} \end{equation}
The row and column of $\mathbf{T}$ are indexed by a combination of two $L$-bit integers $(a,a')$ and $(a',a'')$, so that $N_\mathrm{states} = 2^{2L}$. This construction results in a matrix of size $4^L \times 4^L$. Again we decompose $\mathbf{T}$ into a diagonal matrix $\mathbf{T}^x$ with entries $\mathbf{T}^x_{(a,a'),(a,a')} = e^{-\beta V_x(a,a')}$ and a sparse matrix $\mathbf{T}^z$ with entries $\mathbf{T}^z_{(a,a'),(a',a'')} = e^{-\beta V_z(a, a'')}$ and 0 otherwise. The number of nonzero entries is then $8^L$. The partition function of $N$-layer system is given by $\tr(\mathbf{T}^N)$.
In contrast to Section~\ref{sec:2XAT}, $\mathbf{T}^z$ is no longer symmetric. A general eigensolver is then required to solve the eigenproblem. Although generic eigenvalues can take complex values, the leading eigenvalue is always a positive real number as are the entries of the leading (left and right) eigenvectors, from the Perron-Frobenius theorem. 

Because the value of the nonzero entries in $\mathbf{T}^z$ does not depend on $a'$, $\tmat{\parallel,A}{T}^z_{(a,a'),(a',a'')}$ can be mapped into $\tmat{\perp,A}{T}^z_{a,a''}$ with the interaction strength replaced by $\kappa J$ (instead of $-J$ in Eq.~\eqref{eq:Tz1def}). 
Given that the complexity for the matrix-multiplication with $\tmat{\perp,A}{T}^z$ is $\OO(2^L L)$, and that there are $2^L$ operations (for different $a'$) in total, the complexity for the matrix-multiplication with $\tmat{\parallel,A}{T}^z$ is $\OO(4^L L)$. 
Hence, the matrix-multiplication operation on $\tmat{\parallel,A}{T}$ also has a time complexity of $\OO(4^L L)$. 

\subsection{DNNI Model} \label{sec:2DT}
Because the horizontal and vertical directions of the DNNI model are equivalent, a single transfer matrix can be defined. The contribution of intra-layer states,
\begin{equation}
V_x(a) = -J \sum_{i=1}^L s_i s_{i+1} - h \sum_{i=1}^L s_i,
\end{equation}
is independent of $\kappa$, while the energetic contribution of neighboring layers reads
\begin{equation}
V_z(a, a') = -J \sum_{i=1}^L s_i s'_i + \kappa J \left( \sum_{i=1}^L s_i s'_{i-1} + \sum_{i=1}^L s_i s'_{i+1} \right).
\end{equation}
The transfer matrix $\mathbf{T}$ (shorthand for $\tmat{\perp,D}{T}$) can thus be decomposed into intra-layer and inter-layer interactions as in Eq.~\eqref{eq:tmatdecomp}. For the DNNI model, the inter-layer matrix $\mathbf{T}^z$ is also symmetric (and centrosymmetric when $h=0$) with entries $\mathbf{T}^z_{a,a'}=e^{-\beta V_z(a,a')}$.

Again, $\mathbf{T}^z$ can be decomposed to reduce the complexity of the matrix-vector multiplication, but we can no longer use Eq.~\eqref{eq:xa2tzsimplify}. This scheme drops information about $s'_{i-1}$ after computing the inter-layer interaction for $s_{i-1}$, which, although fine for the ANNNI model, for the DNNI model leaves out the interaction between $s_i$ and its diagonal neighbors, $s'_{i-1}$ and $s'_{i+1}$ [Fig.~\ref{fig:tmatpropagation}(a)].
To make up for this loss of information, we introduce an auxiliary spin $t_1 = s'_i$ for spin indexes $i=1...L$ during the propagation. Because periodic boundary conditions require $s_1 = s_{L+1}$, we introduce an additional auxiliary spin for $r_1 = s'_1$. (For the ANNNI model, $s_L$ does not interact with $s'_1$.) The factorization of $\mathbf{T}^z$ is thus
\begin{equation} \label{eq:2DTztmatprop}
\begin{aligned}
\mathbf{T}^z &= \mathbf{S}^{-1} \mathbf{T}^{z,L} \mathbf{T}^{z,L-1} ... \mathbf{T}^{z,1} \mathbf{S} \\
&= \mathbf{S}^{-1} (\mathbf{P} \mathbf{T}^{z,1|r_1}) (\mathbf{P} \mathbf{T}^{z,1})^{L-1} \mathbf{S}.
\end{aligned}
\end{equation}

The auxiliary matrix $\mathbf{S}$ (and $\mathbf{S}^{-1}$) maps (recovers) a vector of dimension $2^L$ to (from) $2^{L+2}$, namely,
\begin{equation} \begin{aligned}
&(\mathbf{S} x)(\{s_1, s_2, ..., s_L, ..., t_1, r_1\}) = \\
&\quad \begin{cases}
x(s_1, ..., s_L), &t_1=s_L \text{ and } r_1=s_1 \\
0, & \text{ otherwise},
\end{cases}
\end{aligned} \end{equation}
and
\begin{equation}
(\mathbf{S}^{-1} y)(a) = \sum_{t_1, r_1} y(a, t_1, r_1).
\end{equation}

The matrix $\mathbf{T}^{z,1}$, which denotes the contribution on the inter-layer interaction for one spin, $s_1$, has entries
\begin{equation} \label{eq:2DTztmatsingle}
\begin{aligned}
&\mathbf{T}^{z,1}(\{s_1, s_2, ..., s_L, t_1, r_1\}, \{s'_1, s'_2, ..., s'_L, t'_1, r'_1\}) = \\
&\begin{cases}
e^{\beta J s_1 s'_1 - \kappa \beta J s_1 (t'_1 + s'_2) }, & s_i = s'_i (i=2...L), \\ & t_1=s'_1, r_1=r'_1, \\
0, & \text{otherwise}.
\end{cases}
\end{aligned}
\end{equation}
The permutation matrix $\mathbf{P}$ shifts spins $\{s_1, ..., s_L\}$ to $\{s_2, ..., s_L, s_1\}$ but does not change auxiliary spins.
Note that the term $\mathbf{T}^{z,1|r_1}$ in Eq.~\eqref{eq:2DTztmatprop} reflects the periodic boundary condition that replaces $s'_2$ by $r_1$.

The overall time complexity of matrix-vector multiplication remains $\OO(2^L L)$. The size of the temporary vector is quadrupled compared to $\perptm$ for the ANNNI model because two auxiliary spins had to be included, but can be halved by tracing $r_1$ in the code (instead of as a vector index) because it is only invoked for operations on $\mathbf{T}^{z,L}$. 

\subsection{BNNNI and 3NNI models} \label{sec:2BT}

Because the BNNNI model can be considered as a special case of the 3NNI model with $\kappa' = 0$, we only need to consider the later. One possible transfer matrix construction thus requires but minimal modification from $\tmat{\parallel,A}{\mathrm{TM}}$, namely including diagonal nearest-neighbor and axial next-nearest-neighbor interactions in the intra-layer part $\mathbf{T}_{(a, a'), (a, a')}$,
\begin{equation} \begin{aligned}
&\tmat{\parallel,3}{T}^x_{(a,a'),(a,a')} \\ &\quad = \exp\{ -\beta [V_{x}(a,a') + V_{x,\mathrm{B}}(a) + V_{x,\mathrm{D}}(a, a') ] \}
\end{aligned} \end{equation}
where $V_{x,\mathrm{B}}$ and $V_{x,\mathrm{D}}$ are missing from the ANNNI model in Eq.~\eqref{eq:yannnipotentials},
\begin{align}
V_{x,\mathrm{B}}(a) &= -\kappa J \netp(a \bitxor \rol(a, 2)), \\
V_{x,\mathrm{D}}(a, a') &= -\kappa' J [ \netp(a \bitxor \rol(a', 1)) \nonumber \\ &\quad+ \netp(a \bitxor \ror(a', 1)) ].
\end{align}
The structure and complexity of the remaining algorithm then remain unchanged.

However, as stated in the main text (Section~\ref{sec:method:tm}), the checkerboard and diagonal striped phases of the BNNNI and 3NNI models are naturally modulated along the diagonals of a square lattice. To study the correlation length of these modulations and to minimize the finite-size disturbances observed in $\tmat{\parallel,3}{T}$, we thus consider a transfer matrix propagated along the diagonal direction, $\tmat{/,3}{T}$, hence generalizing the approach of Ref.~\cite{oitmaa1987finite}. Note that this arrangement requires $L$ to be even.

As in Eq.~\eqref{eq:tmatdecomp}, the resulting transfer matrix can be decomposed into intra-layer and inter-layer contributions. The intra-layer matrix can be computed directly, and the inter-layer matrix $\mathbf{T}^z$ can be decomposed similarly as for the DNNI model (Eq.~\eqref{eq:2DTztmatprop})
\begin{equation} \label{eq:2BTztmatprop}
\begin{aligned}
\mathbf{T}^z &= \mathbf{S}^{-1} \mathbf{T}^{z,L} \mathbf{T}^{z,L-1} ... \mathbf{T}^{z,2} \mathbf{S} \\
&= \mathbf{S}^{-1} (\mathbf{P} \mathbf{T}^{z,2|t_1, t_2}) (\mathbf{P} \mathbf{T}^{z,2})^{L/2-1} \mathbf{S}.
\end{aligned}
\end{equation}
The auxiliary matrix $\mathbf{S}$ (and $\mathbf{S}^{-1}$) then maps (recovers) a vector of dimension $2^L$ to (from) $2^{L+4}$, 
\begin{equation} \begin{aligned}
&(\mathbf{S} x)(\{s_1, s_2, ..., s_L, ..., t_1, t_2, r_1, r_2\}) = \\
&\quad \begin{cases}
x(s_1, ..., s_L), &(t_1, t_2, r_1, r_2) = (s_{L-1}, s_L, s_1, s_2) \\
0, & \text{ otherwise},
\end{cases}
\end{aligned} \end{equation}
and
\begin{equation}
(\mathbf{S}^{-1} y)(a) = \sum_{t_1, t_2, r_1, r_2} y(a, t_1, t_2, r_1, r_2).
\end{equation}

The matrix $\mathbf{T}^{z,2}$, which denotes the contribution on the inter-layer interaction for two spins, $s_1$ and $s_2$, [Fig.~\ref{fig:tmatpropagation}(c)], has entries
\begin{equation} \label{eq:2BTztmatsingle}
\begin{aligned}
&\mathbf{T}^{z,2}(\{s_1, ..., s_L, t_1, t_2, r_1, r_2\}, \{s'_1, ..., s'_L, t'_1, t'_2, r'_1, r'_2\}) = \\
&\begin{cases}
\exp\{-\beta J [ (s'_1 (t_2 + s_2) & s_i = s'_i (i=3...L), \\
-\kappa (s'_1 (t_1 + s_3) + s'_2 (t_2 + s_4) ) & (t_1, t_2, r_1, r_2) \\
-\kappa' (s'_1 s_1 + s'_2 s_2) ] & =(s'_1, s'_2, r'_1, r'_2), \\
& \\
0, & \text{otherwise}. 
\end{cases}
\end{aligned}
\end{equation}
The permutation matrix $\mathbf{P}$ shifts layer configurations by two spins, i.e., $\{s_1, s_2, ..., s_L\}\rightarrow\{s_3, ..., s_L, s_1, s_2\}$. The term $\mathbf{T}^{z,2|r_1, r_2})$ in Eq.~\eqref{eq:2BTztmatprop} reflects the periodic boundary condition that replaces $s'_3, s'_4$ by $r_1, r_2$. 

This arrangement of auxiliary spins can be viewed as a generalization of the approach used for the DNNI model. A spin $s_i$ here involves interactions with $s_{i \pm 2}$, thus going beyond $s_{i \pm 1}$ for the DNNI model.
In general, for models with inter-layer interactions between $s_i$ and $s'_{i \pm b}$, $2 b$ auxiliary spins are needed, among which $t$ spins are associated with an extended vector, then of size $2^{L+b}$. The size of this vector controls the space complexity for the matrix-vector multiplication. In addition to the $2^b$ loops for different choices of $r$ spins, the time complexity is then $\OO(2^{L+2 b} L)$. 
For $\tmat{/,3}{T}$, in particular, the number of operations and the intermediate vector size are $2^4=16$ and $2^2=4$ times that for $\tmat{\perp,\mathrm{A}}{\mathrm{TM}}$, respectively. 

\section{Reducing space complexity with symmetry} \label{appd:reducedtmat}

In this Appendix we describe computational schemes used to reduce the size of the transfer matrix, and thus significantly decrease the algorithmic space complexity. The key idea is to identify equivalent states in order to construct orthogonal bases (or irreducible representations~\cite{pesch1985transfer}, as have been implemented in related models~\cite{blote1982critical,jin2013phase}). We here adapt this method following the framework of the structured matrix decomposition described in Appendix~\ref{appd:construction}. We first derive the general method for structured matrices with certain permutation invariance, and then analyze the complexity of the transfer matrix involved in solving the models of interest. 

\subsection{General case}

Denote $\mathbf{T}$ as an $n \times n$ matrix that is invariant under permutations $\mathbf{P}_1$, $\mathbf{P}_2$, ..., $\mathbf{P}_g$, such that $\mathbf{P}_i^T \mathbf{T} \mathbf{P}_i = \mathbf{T}$ for $i=1,...,g$ and that these transformations form a symmetry group $\mathcal{G}$. Matrix indexes are then grouped by these transformations. For example, for a centrosymmetric matrix, under the transformation of $\mathbf{P}^T \mathbf{T} \mathbf{P}$ where $\mathbf{P} = (e_n, e_{n-1}, ..., e_1)$, row and column indexes are permuted as $0 \rightarrow n-1$, $1 \rightarrow n-2$ and so on. The indexes $i$ and $n-i-1$ are then deemed equivalent. There are $n/2$ equivalent sets in total.

We next construct a (non-square) matrix with orthogonal columns
\begin{equation} 
\mathbf{Q} \equiv \begin{bmatrix}
1/\sqrt{g_1} & 0 & ... & ... & 0 \\
0 & ... & 1/\sqrt{g_i} & ... & ... \\
... & ... & ... & ... & 1/\sqrt{g_n} \\
1/\sqrt{g_1} & ... & ... & ... & ... \\
\end{bmatrix}
\end{equation}
of size $n \times m$, where $m$ is the number of sets of equivalent indexes (not to be confused with the magnetization). Each column in $\mathbf{Q}$ is a column vector corresponding to an equivalent index set of size $g_i$. The entries in $\mathbf{Q}_i$ are nonzero, and set to $1/\sqrt{g_i}$, if and only if its row index is in the set. Applying the similarity transformation on $\mathbf{T}$ with these bases, we obtain a matrix of size $m \times m$
\begin{equation} \label{eq:mdef}
 \mathbf{M}= \mathbf{Q}^T \mathbf{T} \mathbf{Q}.
\end{equation}

The eigenvalues of $\mathbf{M}$ are also eigenvalues of $\mathbf{T}$, and specifically, $\mathbf{M}$ and $\mathbf{T}$ have the same leading eigenvalue with eigenvector $\varphi_1(\mathbf{T}) = \mathbf{Q} \psi_1(\mathbf{M})$.
 
\begin{figure}
\includegraphics[width=0.95\columnwidth]{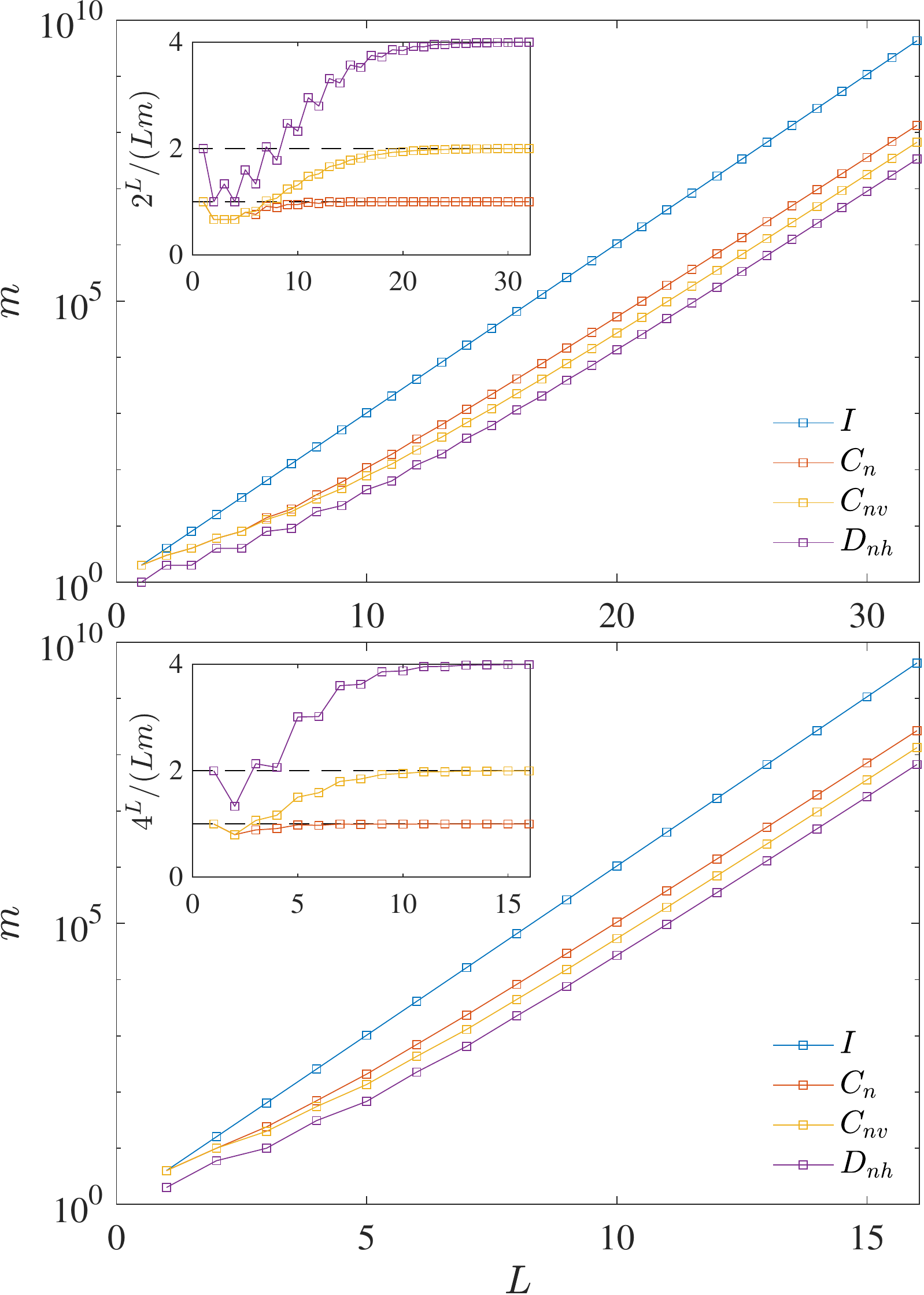}
\caption{Number of equivalent state sets ($m$) of different symmetry group grows with layer length ($L$) in (left) $\perptm$ and (right) $\paralleltm$. Values are obtained in accordance with certain integer series~\cite{noteoeis}. Insets: $m$ asymptotically scales with $N_\mathrm{states}/L$, $N_\mathrm{states}/(2L)$ and $N_\mathrm{states}/(4L)$ under $C_n$, $C_{nv}$ and $D_{nh}$ symmetry, respectively.}
\label{fig:tm-irrep}
\end{figure}

Two-dimensional spin models under periodic boundary condition are invariant under rotation of one spin (a $C_n$ axis) as well as under counting spins backwards
(a $\sigma_v$ reflection), and hence belongs to the $C_{nv}$ point group. In absence of external field, $h=0$, the model is also invariant under flipping all spins (a $\sigma_h$ reflection), and hence it belongs to the $D_{nh}$ group. The dimension of $\mathbf{M}$ is asymptotically reduced by a factor which equals the order of the symmetry group---$2L$ for $C_{nv}$ and $4L$ for $D_{nh}$ (Fig.~\ref{fig:tm-irrep}). In this way, the transfer matrix size can be compressed by a factor of $2L$ (or $4L$ if $h=0$ and only the leading eigenvalue is needed).

\subsection{$\perptm$ and $\diagtm$}

We now adapt this compressed matrix to our transfer matrix calculation. We first consider $\perptm$ (the construction of $\diagtm$ is very similar to $\perptm$, as we will see later). Directly implementing Eq.~\eqref{eq:matvecannni2d} to calculate the matrix-vector multiplication 
\begin{equation} \label{eq:NNreducecomp} 
\begin{aligned}
 \mathbf{w} &= \mathbf{M} \mathbf{v} = \mathbf{Q}^T \mathbf{T} \mathbf{Q} \mathbf{v} = \mathbf{Q}^T (\mathbf{T}^x)^{1/2} (\mathbf{P} \mathbf{T}^{z,1})^L (\mathbf{T}^x)^{1/2} \mathbf{Q} \mathbf{v} \\
 &= \mathbf{Q}^T \mathbf{w}'
 \end{aligned}
\end{equation}
gives essentially the same time and space complexity as for conducting $\mathbf{T} \mathbf{v}$. 
Further optimization is, however, possible. 
Observing that entries in the intermediate vector ($\mathbf{w}'_a$) belonging to the same equivalent states ($a \in b$) are identical, we only need to compute one (among $g_b$ identical entries) for each set of equivalent states in $\mathbf{w}'$ to construct $\mathbf{w}$, such that
\begin{equation}
 \mathbf{w}_b = \sqrt{g_b} \mathbf{w}'_a.
\end{equation}
To take advantage of this property, we initialize an array of equivalent states, denoted $[b]$, containing one of the states ($a$) in the set as well as the set size $(g_b)$. The number of equivalent sets approaches $2^L/(4 L)$ for $h=0$ and $2^L/(2 L)$ otherwise. In both cases $[b]$ has a space complexity of $\OO( 2^L / L)$.
This list can be constructed in two ways with offer different balances in space/time complexity. The first is to set up a temporary array of $2^L$ bits (thus with a $\OO(2^L)$ space complexity) and scanning once (thus with a $\OO(2^L)$ time complexity). The second is to enumerate each of the $2^L$ states and check all of its equivalent states by bit-wise operations (in a $\OO(2^L L)$ time complexity), and push the state to the array only if it has smallest index of all equivalent states. The total number of equivalent state [$\OO(2^L/L)$] then gives the space complexity.

Equation~\eqref{eq:NNreducecomp} can then be evaluated in two parts. First, we compute the inter-layer interactions of all states ${a}$ with $L'$ leftmost spins having the same configuration, $a^\ell$. Denoting the remaining $L-L'$ bits to their right as $a^r$, we hence have $a \equiv a^\ell . a^r$ where ``.'' is a bit concatenation operation. In practice, we set up an intermediate vector $\prescript{a^\ell}{}{\mathbf{v}}$ of size $2^{L-L'}$ such that
\begin{equation} \label{eq:expandv}
\left( (\prescript{0}{}{\mathbf{v}})^T, (\prescript{1}{}{\mathbf{v}})^T, ..., (\prescript{2^{L'}-1}{}{\mathbf{v}})^T \right)^T = \mathbf{Q} \mathbf{v}.
\end{equation}
Decoding each $\prescript{a^\ell}{}{\mathbf{v}}$ from $[b]$ and $\mathbf{v}$ costs a time $\OO(L \times m) = O(2^L)$ and it is run $2^{L'}$ times.
For each $\prescript{a^\ell}{}{\mathbf{v}}$, we further compute
\begin{equation} \label{eq:NNreduceright}
\prescript{a^\ell}{}{\mathbf{w}''} = \prescript{a^\ell}{}{\left((\mathbf{P} \mathbf{T}^{z,1})^{L-L'} (\mathbf{T}^x)^{1/2})\right)} \cdot \prescript{a^\ell}{}{\mathbf{v}}.
\end{equation}
The time complexity of this step is $\OO( (L-L') 2^{L-L'})$ and it is run $2^{L'}$ times (or $2^{L'-1}$ times for $h=0$, because $\mathbf{T}$ is then centrosymmetric). Second, for every non-equivalent entry in $\mathbf{w}'$, we also decompose the index $a' = (a'^\ell . a'^{r})$ and increment $\mathbf{w}'_{a'}$ by $\exp(-\beta V_z(a^\ell, a'^\ell )) \cdot \prescript{a^\ell}{}{\mathbf{w}''_{a'^r}}$. In summary this approach gives
\begin{equation} \label{eq:NNreduceleft}
\mathbf{w}'_{a'} = (\mathbf{T}^x_{a', a'})^{1/2} \sum_{a^\ell} \exp(-\beta V_z(a^\ell, a'^\ell )) \cdot \prescript{a^\ell}{}{\mathbf{w}''_{a'^r}},
\end{equation} 
and the time complexity for this step is $\OO(2^{L'} \times m) = O(2^{L' + L} / L)$.
Comparing the time complexities of Eq.~\eqref{eq:expandv},~\eqref{eq:NNreduceright} and~\eqref{eq:NNreduceleft}, we choose $L' = 1 + \floor{\log_2 L}$ so that the total time complexity for matrix-vector multiplication remains $\OO(2^L L)$ and the space complexity is reduced to $\OO(2^L/L)$. 

The permutation operations that generate equivalent states for $\diagtm$ slightly differs from $\perptm$ because now the layer has a zigzag shape. Specifically, the system is invariant after shifting two (instead of one) spins as well as by first shifting one spin and then counting backwards (instead of simply counting backwards). The number of equivalent states $m$ then asymptotically approaches $2^L/L$ (or $2^L/(2 L)$ for $h=0$). Because we consider two spins $(s_1, s_2)$ in every operation in $\mathbf{T}^{z,2}$, we choose (an even) $L' = 2(1 + \floor{\log_2 {L/2}} )$. The time and space complexities remain the same with $\perptm$, but with a larger prefactor.

\subsection{$\paralleltm$}

Similarly to Eq.~\eqref{eq:NNreducecomp}, for $\paralleltm$ the matrix-vector multiplication is decomposed as
\begin{equation} \label{eq:NNNreducecomp}
 \mathbf{w} = \mathbf{M} \mathbf{v} = \mathbf{Q}^T \mathbf{T} \mathbf{Q} \mathbf{v} = \mathbf{Q}^T (\mathbf{T}^x) (\mathbf{P} \mathbf{T}^{z,1})^L \mathbf{Q} \mathbf{v} = \mathbf{Q}^T \mathbf{w}'.
\end{equation}
Now, however, $\mathbf{M}$ is not symmetric.
In addition to permutation invariance, we can also take advantage of the sparsity of $\paralleltm$ (Sec.~\ref{sec:2YAT}) to compute $\mathbf{M} \cdot \mathbf{v}$. The extra space needed is a vector of size $2^L$ which is much smaller than the vector size of $m=O(4^L/L)$. The time complexity can also be reduced because $\mathbf{w}'$ has identical entries for equivalent states.

The algorithm is as follows. First, we initialize the array of equivalent states $[b]$, as we did for $\perptm$, such that each element is a pair of $L$-bit integers $(a, a')$ that represents this set. Two extra arrays are stored for later bookkeeping purposes: 
\begin{enumerate}
\item An array of equivalent states for $a$ alone, denoted $[c]$, along with the period of $a$ under cyclic shift.
The size of $[c]$ approaches $m' \approx 2^L/(4 L)$ for $D_{nh}$ and $m' \approx 2^L/(2 L)$ for $C_{nv}$ bases.
\item An array of indexes $[b']$ for each representing state $(a, a')$ in $[b]$ that records the references in $[b]$ corresponding to the equivalent state with layers swapped, $(a', a)$, denoted $b'(a', a)$. 
\end{enumerate}
The construction of $[b']$ can follow the construction of $[b]$. For each $(a, a')$ newly appended to $[b]$, we find $b'(a', a)$. If $b'(a', a) \le (a, a')$, a binary search finds the index of $b'(a', a)$ in $[b]$. (Each binary search takes on average $\OO(\ln m) = \OO(L)$ operations, and hence the overall algorithmic complexity remains unchanged.)
In summary, the initialization takes $\OO(4^L)$ operations, and storing $[b]$ takes space $\OO(4^L/L)$.

Second, we setup the subroutine for the matrix-vector multiplication of Eq.~\eqref{eq:NNreducecomp}. Again we denote the indexes of $\mathbf{w}$ and $\mathbf{v}$ as $(a, a')$ and $(a', a'')$, respectively. For each $a'$ in $[c]$, we construct an intermediate vector $\prescript{a'}{}{\mathbf{v}}$ such that
\begin{equation} %\label{eq:expandv2}
\left( (\prescript{a'_1}{}{\mathbf{v}})^T, (\prescript{a'_2}{}{\mathbf{v}})^T, ..., (\prescript{a'_{m'}}{}{\mathbf{v}})^T \right)^T = \mathbf{Q} \mathbf{v}.
\end{equation}
Because each $\prescript{a'}{}{\mathbf{v}}$ is of size $2^L$ and $m'$ of these vectors in total, the complexity of this step is $\OO(2^L m') = O(4^L/L)$.

For each $\prescript{a'}{}{\mathbf{v}}$, we compute
\begin{equation} \label{eq:NNNreduceright}
\prescript{a'}{}{\mathbf{w}''} = \left(\mathbf{P} \cdot \prescript{}{\mathrm{\perp}}{\mathbf{T}^{z,1}}\right)^{L} \cdot \prescript{a'}{}{\mathbf{v}},
\end{equation}
with $J' = -\kappa J$ in $\mathbf{T}^{z,1}$. It takes $\OO( 2^L L)$ operations per $a'$, and hence the total complexity of this step is $\OO(2^L L \times 2^L/L) = \OO(4^L)$.

The entries in the resulting vector, $\prescript{a'}{}{\mathbf{w}''}_a$ corresponds to those of the intermediate vector $\mathbf{w}'$ with
\begin{equation} \label{eq:NNNreduceleft}
\mathbf{w}'_{b'(a, a')} = \mathbf{T}^{x}_{b'(a, a'), b'(a, a')} \prescript{a'}{}{\mathbf{w}''}_a.
\end{equation}
Again, we have
\begin{equation}
\mathbf{w}_b = \sqrt{g_b} \mathbf{w}'_a,
\end{equation}
and hence Eq.~\eqref{eq:NNNreduceleft} needs to be evaluated $m = O(4^L / L)$ times.

In summary, the time complexity is of $\OO(4^L)$, an improvement by a factor of $L$ over that in Sec.~\ref{sec:2YAT}. The extra space needed is $\OO(m')$ which is marginal given that input and output vectors have sizes $\OO(m)$.

\subsection{Remarks}

Thanks to these compressed TMs, evaluation of systems with $L$ up to 36 for $\perptm$, 32 for $\diagtm$ and 16 for $\paralleltm$ is accessible within $~60$ GB memory.
Interestingly, when evaluating $\mathbf{T}$ with an iterative eigensolver, convergence slows down markedly around transition temperatures. In $\perptm$ for the ANNNI model with $L=20$, for example, the slowdown at $T_\mathrm{c}$ can be as much as $15\times$ that of a typical run away from that temperature. The slowdown for the compressed TM, however, is less pronounced, which facilitates free energy calculations. The compressed TM is therefore better conditioned, which adds extra advantage to this consideration. As a result, the algorithm is numerical stable and generates results with high accuracy. This property is indeed related to the underlying physics. The condition number is defined as the error of the output given an erroneous (or finite precision) input. The specific heat also corresponds to the fluctuation on energy, $\langle (\delta u)^2 \rangle = k_B T c$, which means when $c$ is high, the output of the eigenvectors (density of configurations) is very unstable, and leads to a large condition number. The phase transition in physics and convergence theory in computer science is intrinsically related.

A potential challenge for this decomposition, however, is that subleading eigenvalues of $\mathbf{T}$ can lie either in $\mathrm{span}(\mathbf{M})$ or in $\mathrm{null}(\mathbf{M})$.
In other words, the spectral gap, which gives the leading correlation length, of $\mathbf{M}$ does not necessarily coincide with that of $\mathbf{T}$. 
In practice, it is observed that for $\tmat{\parallel,A}{\mathbf{T}}$, $\tmat{\perp,D}{\mathbf{T}}$, $\tmat{\parallel,3}{\mathbf{T}}$ and $\tmat{/,3}{\mathbf{T}}$ in $h=0$, the subleading eigenvectors are all skew-symmetric, and preserved in the compressed TMs with $C_{nv}$ bases. We therefore identify the correlation length from the compressed matrix. For $\tmat{\perp,A}{\mathbf{T}}$, however, the subleading eigenvalue is doubly degenerate (as in Ref.~\onlinecite{pesch1985transfer}), and is observed in the null space of $\mathbf{M}$ obtained from $C_{nv}$ bases. The original transfer matrix is thus used to identify the correlation length. 
As noted in Ref.~\onlinecite{pesch1985transfer} this is a result of symmetry of the bases. For other models a similar argument might also be possible.

% references
\bibliography{abbrev}

\end{document}